\documentclass[twocolumn,pra,aps,showpacs,superscriptaddress,footinbib]{revtex4-1}
\usepackage[english]{babel}
\usepackage[utf8]{inputenc}

\usepackage{url,psfrag,graphicx}
\usepackage{dcolumn}
\usepackage{amsmath,amssymb,amsthm}
\usepackage{bm}
\usepackage{pstricks}
\usepackage{hyperref}
\usepackage{epsfig}
\usepackage{verbatim}
\def\<{\left<}
\def\>{\right>}
\def\ket|#1>{\left|#1\right>}
\def\bra<#1|{\left<#1\right|}

\def\elem<#1|#2|#3>{\left<#1\right|#2\left|#3\right>}
\def\({\left(}
\def\){\right)}

\def\J{{\cal J}}
\def\O{{\cal O}}
\def\A{{\cal A}}
\def\all{\text{all}}
\def\av{\text{av}}

\usepackage{graphicx}
\begin{document}

\title{Random walkers on a deformable medium}

\author{Carlos Lajusticia-Costan}
\affiliation{Departamento de Física Fundamental, Universidad Nacional de
  Educación a Distancia (UNED), Madrid, Spain}
\author{Silvia N. Santalla}
\affiliation{Departamento de Física and Grupo Interdisciplinar de
  Sistemas Complejos (GISC), Universidad Carlos III de Madrid,
  Madrid, Spain}
\author{Javier Rodríguez-Laguna}
\affiliation{Departamento de Física Fundamental, Universidad Nacional de
  Educación a Distancia (UNED), Madrid, Spain}
\author{Elka Korutcheva}
\affiliation{Departamento de Física Fundamental, Universidad Nacional de
  Educación a Distancia (UNED), Madrid, Spain}
\affiliation{G. Nadjakov Institute of Solid State Physics, Bulgarian
  Academy of Sciences, 1784 Sofia, Bulgaria} 

\begin{abstract}
We consider random walkers that deform the medium as they move,
enabling a faster motion in regions which have been recently
visited. This induces an effective attraction between walkers mediated
by the medium, which can be regarded as a space metric, giving rise to
a statistical mechanics toy model either for gravity, motion through
deformable matter or adaptable geometry. In the strong-deformability
regime, we find that diffusion is initially described by the {\em
  porous medium equation}, thus yielding subdiffusive behavior of an
initially localized cloud of particles. Indeed, while the average
width of a single cloud will sustain a $\sigma\sim t^{1/2}$ growth,
the combined width of the whole ensemble will grow like $\sigma\sim
t^{1/3}$ in a certain time regime. This difference can be accounted
for by the strong correlations between the particles, which we explore
indirectly through the fluctuations of the center of mass of the cloud
and the expected value of the experienced density, defined as the
average density measured by the particles themselves.
\end{abstract}

\date{May 6, 2021}

\maketitle


\section{Introduction}

Let us consider random walkers in a forest leaving some trail behind
them, which they may come across again at a later time. The walkers
may then choose to avoid that trail or, otherwise, be attracted
towards it. Thus, the walkers interact with the medium, which will
interact with the walkers in its turn. Alternatively, we may consider
the walkers to have a memory of past events, changing their behavior
appropriately.

Random walkers with a memory, which are also called non-Markovian,
have been described thoroughly. Specially interesting is the {\em
  elephant random walk} (ERW), in which the random walker copies the
step taken some random time ago
\cite{note_erw,schutz.04,kumar.10,dasilva.14,kearney.18}. Random
walkers can be interpreted as foragers gathering food, and in that
case they may decide from time to time to jump to a previous point in
their search path, where they were specially successful. This gives
rise to the {\em random walks with relocation}
\cite{boyer.14,boyer.14.2,boyer.16}, which are very useful to describe
animal motion, and to develop search algorithms
\cite{campos.15,falcon.17}.

{\em Self-avoiding walks}, on the other hand, were developed in the
context of polymer science \cite{degennes.79}. The walk is assigned a
probability globally, not step-by-step, which becomes zero when the
walk intersects itself \cite{amit.83}. Alternatively, random walkers
can be either forbidden to cross their previous path, thus allowing
them to become trapped, or they may receive an energy penalty for each
self-cross \cite{domb.72,stanley.83}.

Yet, random walkers may present an affinity towards already visited
sites, as it is the case with {\em taxis} in biology
\cite{othmer.97,permantle.07}, forming {\em reinforced random-walks}
where the walker finds it more likely to step on an already visited
site, with memory sometimes fading with time, often to the extreme of
only remembering the last step
\cite{davis.90,sapozhnikov.94,ordemann.01,foster.09}. Random walkers
may also create a potential well around the visited sites, thus
energetically encouraging future visits
\cite{huang.02,tan.02}. Moreover, they may interact by leaving a
tracing chemical substance, in the process known as {\em chemotaxis}
\cite{Schweitzer.19}. Such reinforced random-walks may present lack of
self-averaging and give rise to interesting open problems, such as the
determination of the trapping probability \cite{kozma.12}. Similar
ideas have been successfully applied to the design of optimization
algorithms on complex landscapes, such as the {\em ant colony
  optimization}\cite{ant_colony} or the {\em river formation}
heuristics\cite{river_formation}.

In this work we consider random walkers that reduce the {\em waiting
  time} associated with a link each time they traverse it
\cite{miyazama.87,bouchaud.90,li.17}. This can be considered
equivalent to a change in the local speed of light or, in other words,
a deformation in the space-time metric. Thus, walkers move faster in
regions which have been recently visited. Yet, we will consider the
metric to present a tendency to return to its undeformed state when
the walkers are not disturbing it. Interestingly, the random walkers
will feel an effective attraction mediated by the metric deformations,
giving rise to strongly correlated clusters which suggest that the
system may serve as a toy model for stochastic gravity \cite{hu.04}.

In the continuum limit, the position-dependent waiting times map to an
inhomogeneous diffusivity. Thus, we can write a generalized
diffusion equation with two fields: {\em matter} is described by
the probability distribution for the walker positions, and the {\em
  medium} or {\em geometry} is described by the diffusivity field. The
associated partial differential equations are non-linear, and in the
strong coupling limit we will show that they are ruled by the {\em
  porous medium equation} \cite{vazquez.91}, with non-trivial scaling
exponents showing subdiffusive behavior.

This article is organized as follows. Section \ref{sec:model}
describes our discrete 1D random-walk model on a deformable medium
(RWDM). The continuum limit of this model is considered in
Sec. \ref{sec:continuum}, which discusses the expected scaling
properties. Extensive numerical simulations are performed and
discussed in Sec. \ref{sec:simul}, which give rise to a physical
picture which is discussed in Sec. \ref{sec:picture}.
Sec. \ref{sec:conclusions} provides a summary of the conclusions and
proposals for further work. 


\begin{figure*}
  \center
  \includegraphics[width=12cm]{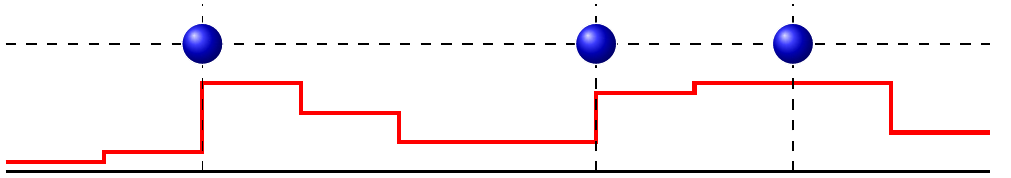}
  \includegraphics[width=16cm]{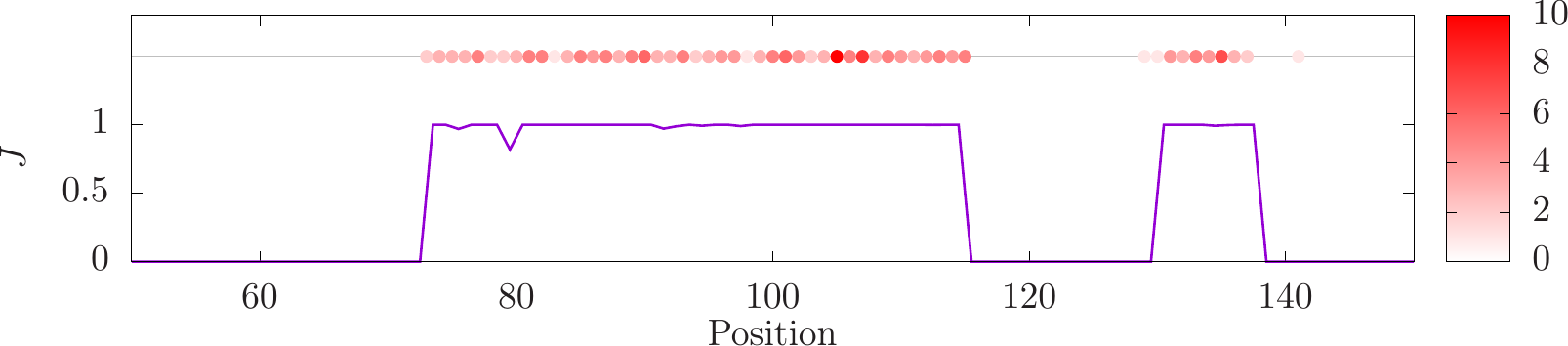}
  \caption{Top: Illustration of the RWDM model. Walkers are
    represented by the balls, while the hopping probabilities between
    neighboring sites are represented by the red line. Bottom: A
    configuration for $\J_0=10^{-6}$, $\J_1=1$, $t_0=10$ and $N_P=200$
    particles on $L=200$ sites, after $T=10^8$ time-steps. The purple
    line denotes the hopping probabilities, red balls represent the
    positions of the particles, and the color intensity denotes the
    number of particles at a certain site. Notice that most hopping
    probabilities are near $\J_0$ or $\J_1$, and that the particles
    are neatly divided into two blocks.}
  \label{fig:illust}
\end{figure*}

\section{Random-Walk on Deformable Medium}
\label{sec:model}

Let us describe our model, which will be called the {\em random walk
  on a deformable medium (RWDM)}. We consider $N_P$ random walkers on
a chain. When a walker is standing on site $i$, it has a certain
probability per unit time $J_{i,i-1}$ of hopping towards site $i-1$,
and $J_{i,i+1}$ of hopping towards site $i+1$. We will assume that
probabilities are symmetric, i.e. $J_{i,i+1}=J_{i+1,i}$, and we will
introduce the convenient notation $J_i\equiv J_{i,i+1}$. The set of
$\{J_p\}$ values will be called the {\em medium} or the {\em
  metric}. Notice that the hopping process on a single link follows a
Poissonian law, and that the expected {\em waiting time} before
hopping is inversely proportional to the hopping rate. Thus, the
walker may take either its left link with hopping rate $J_{i-1}$ or
its right link with hopping rate $J_i$. The top panel of
Fig. \ref{fig:illust} provides an illustration.

The $N_P$ particles are initially placed at the center of a size $L$
chain with periodic boundaries and all hopping probabilities
$J_i=\J_0$ \cite{numerical}. A suitable time-step is chosen, $\Delta
t$. At each turn, we select a random particle and look up its
location, $i$. We decide whether to attempt a jump to the left or to
the right with equal probabilities. The jump attempt will succeed with
probability $P=J_k\Delta t$, where $k=i-1$ for a left jump, or $k=i$
for a right jump. After a successful jump attempt, the link will
increase its hopping probability according to the rule

\begin{equation}
  J_k \to J_k + {\Delta t\over t_0}\(\J_1-J_k\),
  \label{eq:Jincrease}
\end{equation}
where $\J_1\geq \J_0$, thus completing the turn. After a round of
$N_P$ turns, we assert that time has advanced by a time-step $\Delta
t$. Then, all links which have not been updated during this round are
subject to the relaxation rule,

\begin{equation}
  J_k \to J_k - {\Delta t\over t_0}\(J_k-\J_0\).
  \label{eq:Jdecay}
\end{equation}
Notice that, in the continuum limit, rules \eqref{eq:Jincrease} and
\eqref{eq:Jdecay} correspond respectively to an exponential increase
or decrease towards the maximum and minimum values, $\J_1$ and $\J_0$,
with the same characteristic time $t_0$. A typical configuration after
a long simulation time can be seen in the bottom panel of
Fig. \ref{fig:illust}.

If $\J_1=\J_0$, the metric becomes rigid and the particles will follow
a standard random walk. Thus, in this limit we expect that the
particle cloud will behave diffusively, i.e. the expected deviation of
the particle positions will scale like $\sigma\sim t^{1/2}$. If
$\J_1> \J_0$ then the particles will present a trend to hop on
already used links, thus spending more time on the regions already
visited. This may lead to {\em self-localization} of the particle
cloud, through interaction with the medium. In this case, we may still
expect a power-law behavior, with the traversed distance growing like
$t^\alpha$ and $\alpha<1/2$.

Our model can be considered as a type of {\em reinforced random walk
  model}, and as such it may show absence of self-averaging. Yet, our
system presents an important difference. In a typical reinforced
random walk model particles are more likely to jump on sites already
visited \cite{davis.90}, sometimes through an energy advantage
\cite{huang.02,tan.02}. In our model the medium is deformed through a
change in the hopping rates. Sites do not attract particles, instead
they move faster through recently visited regions.

\vspace{1mm}

\begin{figure}
  \includegraphics[width=8.5cm]{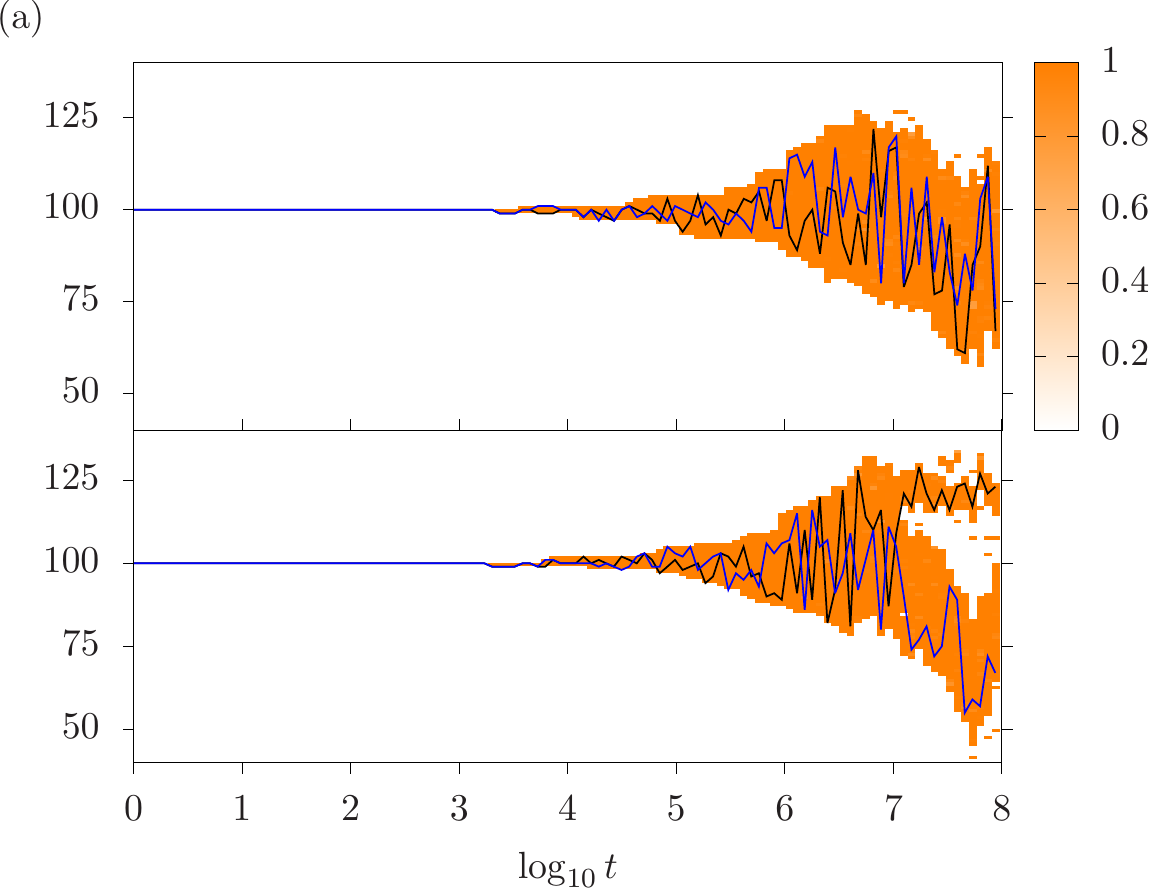}
  \includegraphics[width=8.5cm]{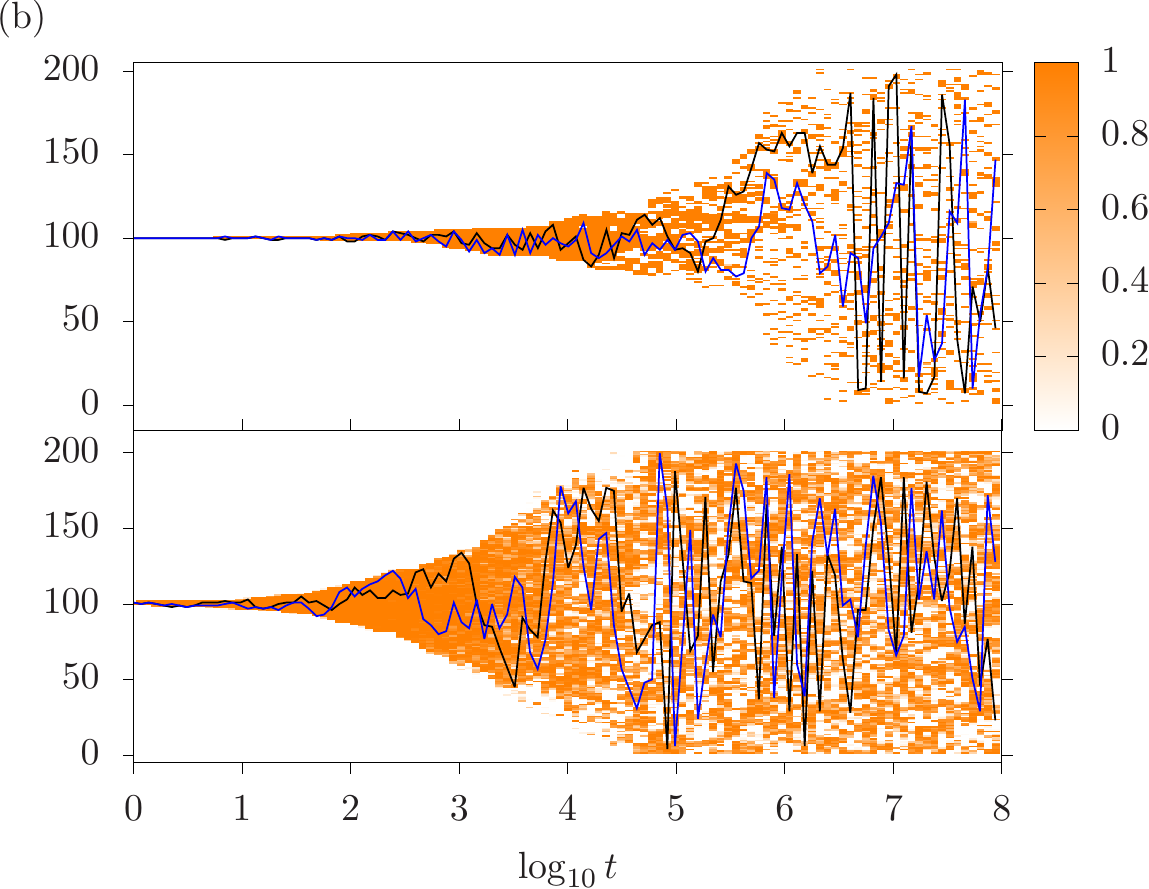}
  \caption{(a) Two system histories, using $N_P=200$ particles,
    $\J_0=10^{-6}$, $\J_1=1$, $t_0=10$ and $L=200$. Time is
    represented in the horizontal axis, in logarithmic scale, while
    the Y-axis stands for positions along the chain. Color denotes the
    local hopping probability, and two particle histories (out of
    $N_P=200$) are depicted with continuous lines. (b) Similar results
    using $\J_0=10^{-4}$ (top) and $\J_0=10^{-2}$ (bottom). Notice
    that larger values of $\J_0$ give rise to sparse distributions of
    the particle cloud.}
  \label{fig:bruja}
\end{figure}

Fig. \ref{fig:bruja} (a) shows two histories, using $\Delta t=1$,
$t_0=10$, $\J_0=10^{-6}$ and $\J_1=1$ on a $L=200$ chain with
$N_P=200$ particles. The X-axis represents time in logarithmic scale,
while the Y-axis represents the position. Color intensity denotes the
hopping probabilities $J_i$, where we can see that most values are
close to either $\J_0$ or $\J_1$. The width of the cloud at a certain
time can be estimated from the vertical range of the high hopping
region, ${\cal H}\equiv \{i | J_i\approx \J_1\}$. Indeed, the cloud
stays localized at the center up to times of order $\approx
3000$. Afterwards, it increases its width for some time, and then
either opens up (top) or drifts laterally (bottom). Two trajectories
of individual particles have been traced, showing that the walkers
spread all over the high-hopping region.

Fig. \ref{fig:bruja} (b) shows the same data for $\J_0=10^{-4}$ (top)
and $\J_0=10^{-2}$ (bottom). Notice that the clouds are much more
sparse, suggesting that the particles do not constitute a compact
cluster in this case.


\section{Mean-field Continuum description}
\label{sec:continuum}

In order to determine the macroscopic behavior of our model, we will
assume that matter can be described through a density function,
$P(x,t)$, such that the expected number of particles in a region
$\A$ will be given by

\begin{equation}
\<N_\A(t)\>=N_P\int_\A dx\; P(x,t).
\label{eq:na}
\end{equation}
where $P(x,t)$ will be assumed to be smooth. Moreover, the hopping
terms will be described by a smooth function too, $J(x,t)$, which we
will call the metric field.

The evolution of the density function will be ruled by a diffusion
equation through an inhomogeneous medium \cite{vazquez.91}

\begin{equation}
\partial_t P(x,t)= \partial_x \Big( J(x,t) \partial_x P(x,t) \Big).
\label{eq:fokkerplanck}
\end{equation}
This equation can be understood as a diffusion equation on a
metric of the form

\begin{equation}
  ds^2 = J^2(x,t)\, \(-dt^2+dx^2\),
  \label{eq:metric}
\end{equation}
which is conformally equivalent to the Minkowski metric,
$ds^2=-dt'^2+dx'^2$, upon a change of coordinates of the form

\begin{align}
dx'&=J(x,t)\,dx, \nonumber\\
dt'&=J(x,t)\,dt.
\label{eq:metricchange}
\end{align}
Moreover, we observe that Eq. \eqref{eq:fokkerplanck} transforms into the
standard heat equation, 

\begin{equation}
  \partial_{t'} P(x',t') = \partial^2_{x'} P(x',t').
  \label{eq:laplacian}
\end{equation}

Yet, the metric field $J(x,t)$ is not fixed beforehand in our
model. Instead, it depends dynamically on the matter field $P(x,t)$,
giving rise to a back-reaction of the matter into the metric. Indeed,
we may claim that the geometry tells matter how to move through
Eq. \eqref{eq:fokkerplanck}, while matter tells geometry how to curve
through an equation that will be deduced next.

Wherever there are moving particles over a link, the hopping
probability increases up to $\J_1$, and where there are no particles,
the hopping probability decays to $\J_0$. Thus, we can propose the
following equation

\begin{align}
  \partial_t J(x,t) & = K_0 P(x,t)\Big[\J_1 - J(x,t)\Big]  \nonumber\\
  + & K_1 \Big(1-P(x,t)\Big)\Big[\J_0 - J(x,t)\Big],
\label{eq:diffusionlinearapproach}
\end{align}
where $K_0$ and $K_1$ are the increasing and decreasing rates,
respectively. In our microscopic model, $K_1=K_0=K \approx 1/t_0$. Thus,
we can reorganize the terms,

\begin{equation}
\partial_t J(x,t) = K\Big[P(x,t)(\J_1-\J_0) + \J_0 - J(x,t)\Big].
\label{eq:diffusiondependency}
\end{equation}

Let us consider the limit of very fast back-reaction between the
metric and the matter fields, $K\to\infty$ or $t_0\to 0$. In that
limit, the term in parenthesis in Eq. \eqref{eq:diffusiondependency}
must vanish, and we have

\begin{equation}
J(x,t) = P(x,t)\Big(\J_1-\J_0\Big) + \J_0,
\label{eq:diffusiondependency3}
\end{equation}
Plugging this equation into our original inhomogeneous diffusion
Eq. \eqref{eq:fokkerplanck}, we can find a non-linear equation
for the matter field alone, 

\begin{align}
  \partial_t P(x,t) &= \Delta\J(\partial_x P(x,t))^2 \nonumber\\
  +& \Big[\J_0 +\Delta\J P(x,t)\Big]\partial^2_x P(x,t),
  \label{eq:diffusiondependency4}
\end{align}
where $\Delta\J\equiv \J_1-\J_0$. We can also solve for $J(x,t)$,
resulting in

\begin{equation}
\partial_t J(x,t)= \partial_x \Big( J(x,t) \partial_x J(x,t)\Big).
\label{eq:diffusiondependency5}
\end{equation}

Furthermore, in the limit in which $\J_0\to 0$ and $\Delta\J\to \J_1$
we reach an effective equation for the matter field,

\begin{align}
  \partial_t P(x,t) & = \J_1 \Big[ (\partial_x P(x,t))^2 +
  P(x,t)\partial^2_x P(x,t) \Big] \nonumber \\
  & = {\J_1\over 2} \partial^2_x \(P^2(x,t)\).
  \label{eq:pme}
\end{align}

\begin{figure}
  \includegraphics[width=\linewidth]{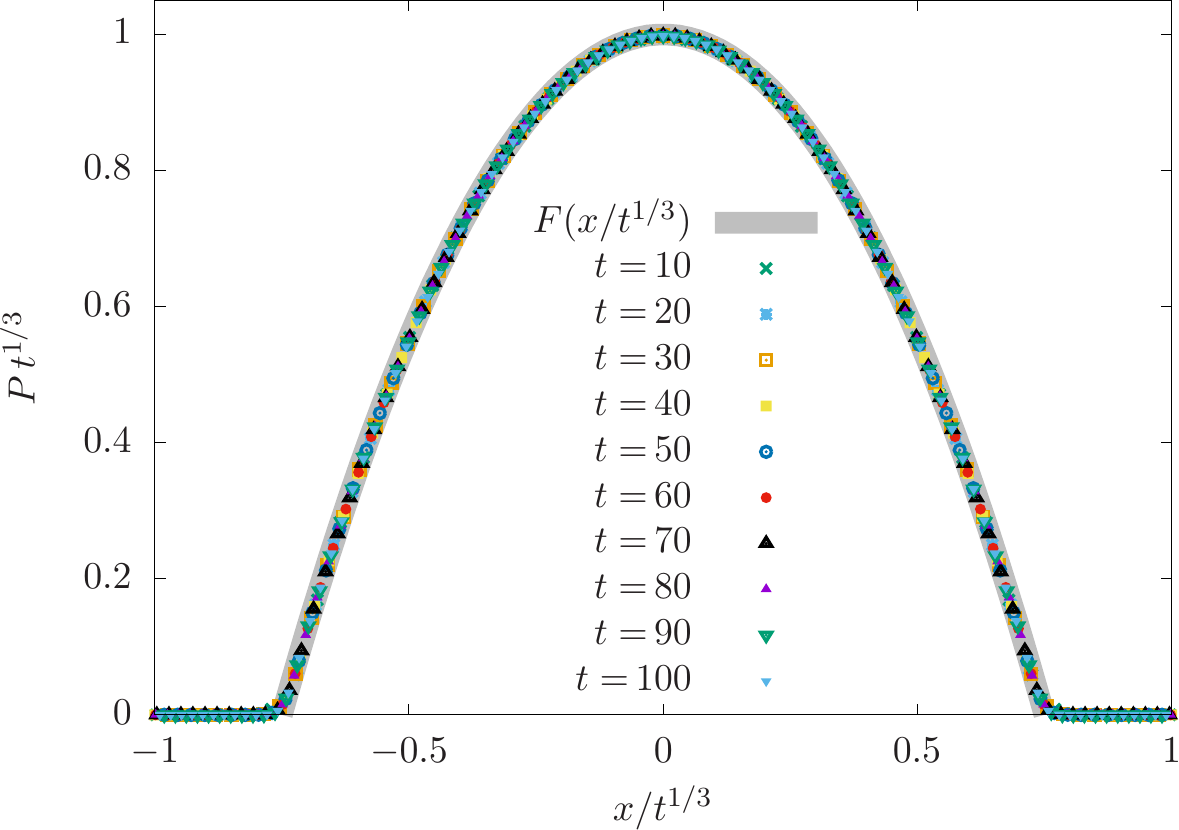}
  \caption{Rescaled solutions of the porous medium equation,
    Eq. \eqref{eq:pme} with a Dirac delta function as the initial
    condition, showing the collapse predicted in
    Eq. \eqref{eq:collapse}.}
 \label{fig:pme}
\end{figure}

This last equation is a parabolic partial differential equation known
as the {\em porous medium equation} (PME) \cite{vazquez.91}, which is
a non-linear relative of the heat equation. Opposite to the heat
equation, the PME is {\em causal}, presenting a well-defined
light-cone. Also, the scaling properties are very different from the
heat equation. Indeed, it has solutions of the form

\begin{equation}
  P(x,t)=t^{-1/3}P\(xt^{-1/3},1\)=t^{-1/3}F\(xt^{-1/3}\),
  \label{eq:collapse}
\end{equation}
where $F$ is a universal function which is provided in Appendix
\ref{sec:appendix} \cite{Pattle.59}. We have checked numerically the
validity of Eq. \eqref{eq:collapse} in Fig. \ref{fig:pme}. In order to
perform the check, we have obtained numerically the solution of the
PME, Eq. \eqref{eq:pme}, with a Dirac delta function as the initial
condition, $P(x,0)=\delta(x)$. The solution is suitably rescaled,
showing $t^{1/3} P(x,t)$ as a function of $xt^{-1/3}$ for several
values of $t$. The different curves collapse into the universal
function $F$, as predicted.

Note that this theoretical framework suggests that the spread of the
particle cloud will grow sub-diffusively, $\sigma \sim t^{1/3}$. 

It is relevant to ask ourselves about the validity regime of the
previous approximation. For example, we have considered the limit
$t_0\to 0$ in the determination of
Eq. \eqref{eq:diffusiondependency3}. This approximation is valid when
$t_0$ is small compared to the time scale required for particles to
jump over a $\J_0$ link, i.e. $\J_0^{-1}$, which will be the case for
all our simulations. Moreover, we assume that the probability
distribution and the hopping rates are smooth enough, which need not
be the case for a single sample, but will be true for the ensemble
average.


\section{Numerical Simulations}
\label{sec:simul}

We have considered an $L=200$ chain with periodic boundaries, using
$\J_1=1$, $t_0=10$ and $\Delta t=1$, similarly to the two histories
shown in Fig. \ref{fig:bruja} (a). In our simulations $N_P=25$, $50$,
100 or 200 particles, and $\J_0=10^{-6}$, $10^{-4}$ or $10^{-2}$. In
all cases, we have simulated up to time $T_{\text{max}}=10^8$, taking
$N_S=2000$ samples (unless otherwise stated). Throughout this section
we define $x_{p,s}(t)$ as the position of particle $p$ in sample $s$
at time $t$, and $J_{i,s}(t)$ as the hopping probability between sites
$i$ and $i+1$ in sample $s$ at time $t$. The initial condition is
given by

\begin{align}
  x_{p,s}(0)=L/2, \qquad & \text{for all $p$ and $s$,}\\
  J_{i,s}(0)=\J_0, \qquad & \text{for all $i$ and $s$.}
  \label{eq:initial}
\end{align}

We must make a distinction between averaging over all particles within
each sample, $E_P[\cdot]$, averaging over all samples, $E_S[\cdot]$,
and averaging over both simultaneously, $E_{SP}[\cdot]$, defined as

\begin{align}
  E_S[q_s] &\equiv {1\over N_S} \sum_{s=1}^{N_S} q_s, \\
  E_P[q_p] &\equiv {1\over N_P} \sum_{p=1}^{N_P} q_p, \\
  E_{SP}[q_{p,s}] &\equiv E_S[E_P[q_{p,s}]] = E_P[E_S[q_{p,s}]],
  \label{eq:mean}
\end{align}
where $q_s$, $q_p$ and $q_{p,s}$ are observables associated to each
sample, each particle or both. Also, we can define the corresponding
variances, $\sigma_\O^2[X]\equiv E_\O[X^2]-E_\O[X]^2$, for
$\O\in\{S,P,SP\}$. 

We have considered the evolution of the average width of the particle
cloud, defined in two alternative ways. First of all, we can consider
the {\em global} deviation of the positions of all particles in all
samples, 

\begin{equation}
  W_{\text{all}}(t)=\sigma_{SP}[x_{p,s}(t)],
  \label{eq:widthall}
\end{equation}
and we can compare it with the average value of the widths of each
sample, which is defined as

\begin{equation}
  W^2_{\text{av}}(t)=E_S[\sigma^2_P[x_{p,s}(t)]].
  \label{eq:widthav}
\end{equation}

Necessarily, the deviation of the whole ensemble of particle positions
must exceed (or be equal to) the average of the deviations within each
sample,

\begin{equation}
  W_{\text{all}}(t)\geq W_{\text{av}}(t).
  \label{eq:ineq}
\end{equation}

\begin{figure}
  \includegraphics[width=8cm]{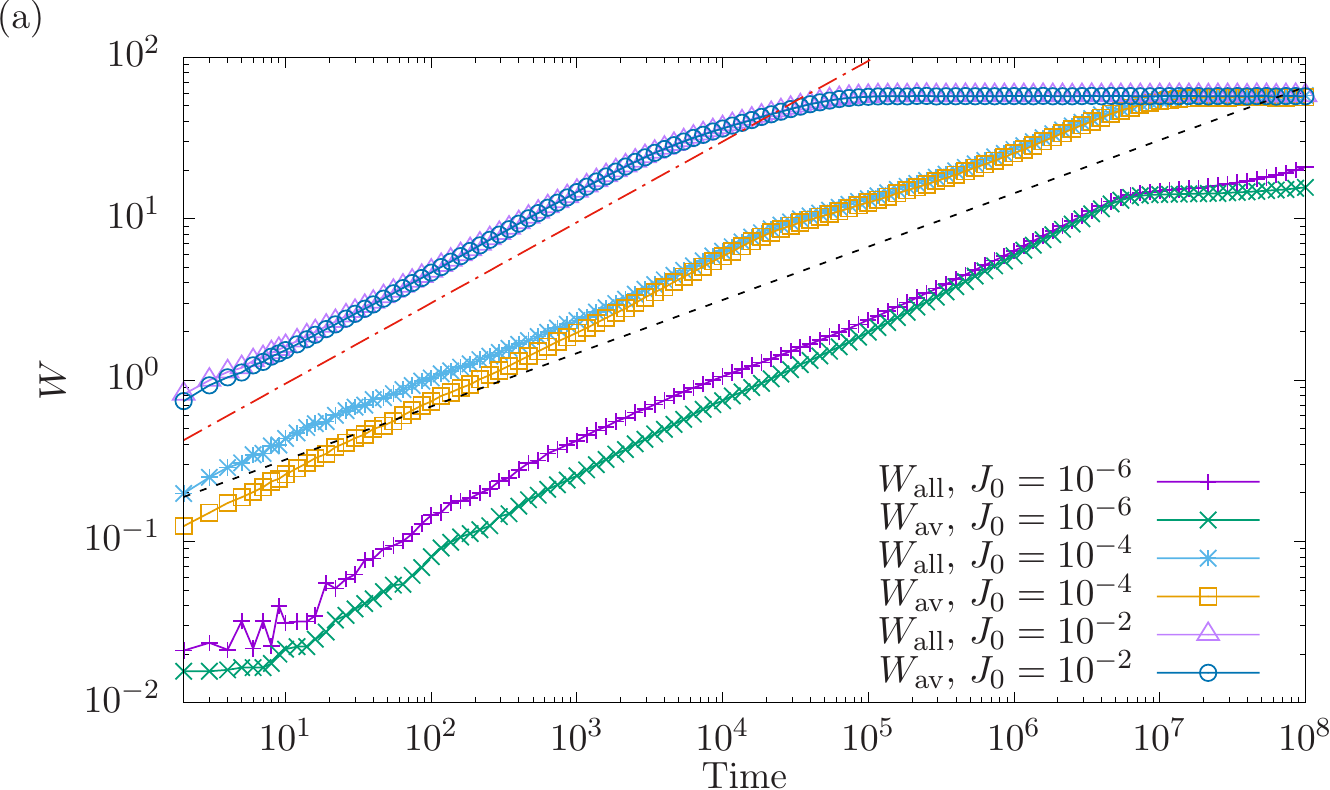}
  \includegraphics[width=8cm]{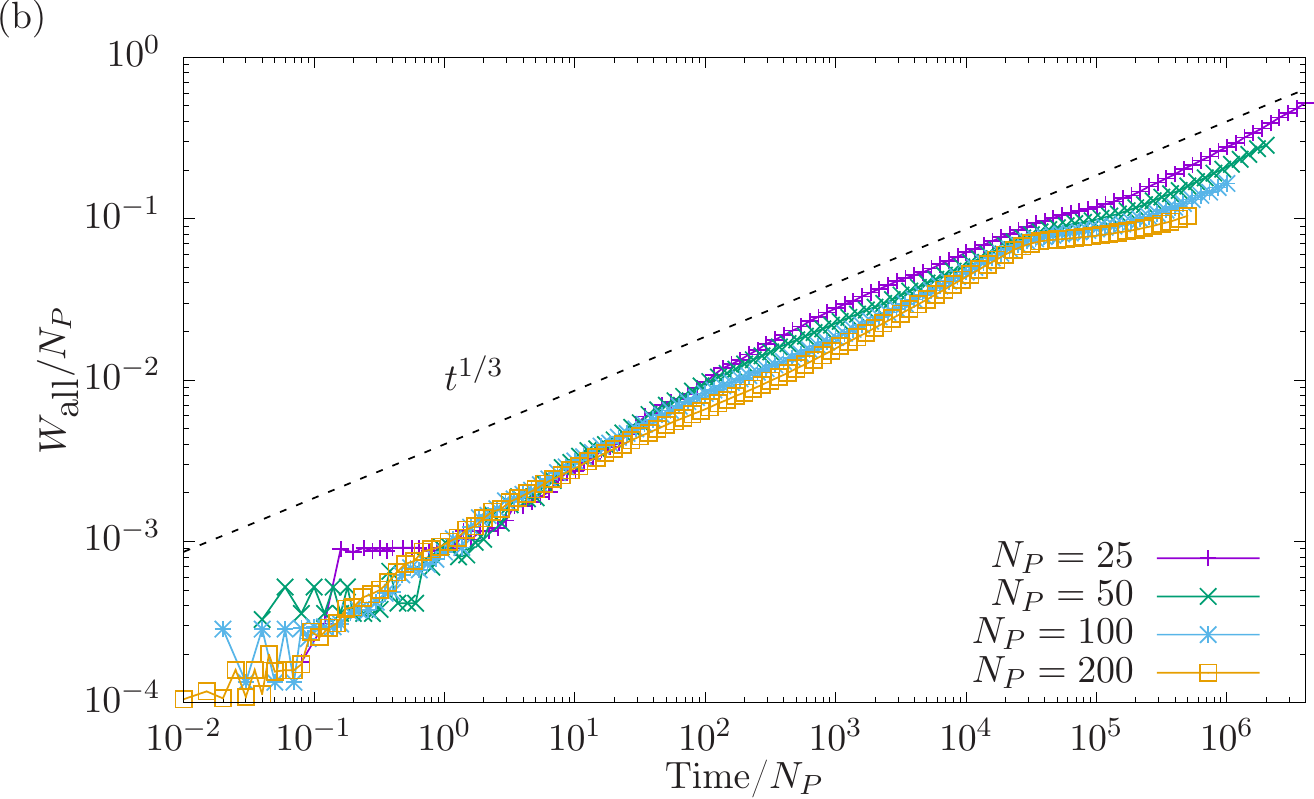}
  \includegraphics[width=8cm]{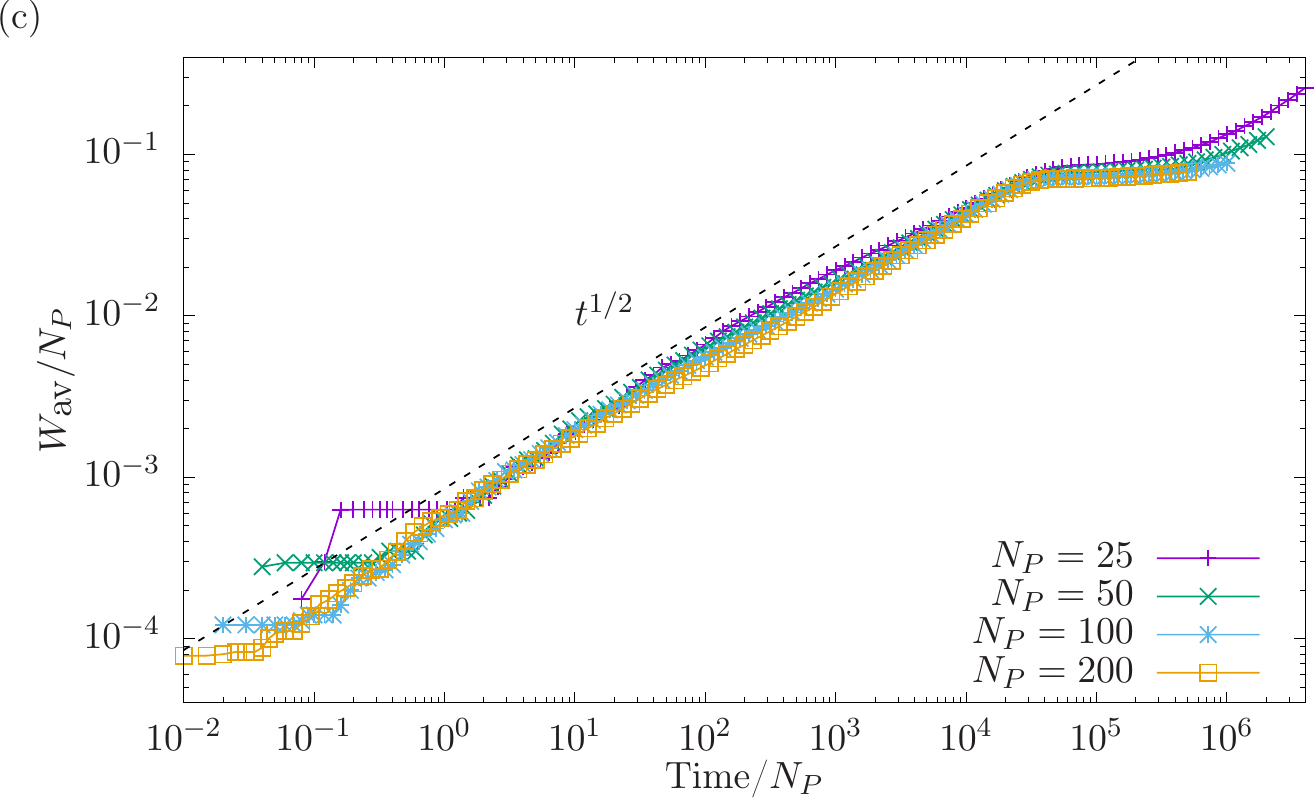}
  \caption{Average width of the particle cloud, as a function of
    time. (a) Comparison of $W_{\text{all}}(t)$,
    Eq. \eqref{eq:widthall}, and $W_{\text{av}}(t)$,
    Eq. \eqref{eq:widthav}, using $N_P=200$ and three different values
    for $\J_0=10^{-6}$, $10^{-4}$ and $10^{-2}$. The top straight line
    shows a $t^{1/2}$ behavior, and the lower one $t^{1/3}$. Notice
    that inequality \eqref{eq:ineq} holds exactly in all
    cases. (b) $W_{\text{all}}$ as a function of time for
    $\J_0=10^{-6}$, both rescaled with the number of particles $N_P$,
    showing the power-law $t^{1/3}$. (c) $W_{\text{av}}$ as a
    function of time for $\J_0=10^{-6}$, rescaled in the same way,
    while the dashed line represents $t^{1/2}$.}
  \label{fig:width}
\end{figure}

Fig. \ref{fig:width} (a) shows both widths using $N_P=200$ and three
different values of $\J_0$. Both widths grow approximately as a
power-law for a broad range of times. For $\J_0=10^{-2}$ both widths
coincide, growing diffusively (dashed-dotted line denotes $t^{1/2}$)
up to a certain saturation time, while for $\J_0=10^{-4}$ we observe a
significant difference between $W_\all$ and $W_\av$ only for short
times (notice that the vertical axis is in logarithmic scale), and a
wide time range in which both grow as $t^{1/3}$. For $\J_0=10^{-6}$, 
on the other hand, we see a more significant difference between both 
widths. After an initial transient time, $W_\av$ grows diffusively
up to times $\sim 10^{7}$ whereas the growth of $W_\all$ is mostly 
subdiffusive, particularly in the interval $t \approx 5\cdot10^5$ to $t 
\approx 10^7$. 
From that moment on, $W_\av$ approximately saturates whereas $W_\all$
keeps growing. The next sections will provide a physical picture for
that complex behavior. Indeed, we contend that $W_\all$ corresponds
approximately to the expected behavior for the width of the particle
cloud predicted in by the continuum approximation for very low $\J_0$,
provided by the scaling of the PME, Eq. \eqref{eq:pme}, as we can see
in Eq. \eqref{eq:collapse}. The reasons for the gap between $W_\all$
and $W_\av$ will be discussed in the next section, and are related to
the absence of self-averaging.

In Fig. \ref{fig:width} (b) and (c) we see that both $W_\all$ (b) and
$W_\av$ (c) collapse approximately for $\J_0=10^{-6}$ when the time
and width axes are rescaled dividing by the number of particles $N_P$.

\begin{figure}
  \includegraphics[width=8cm]{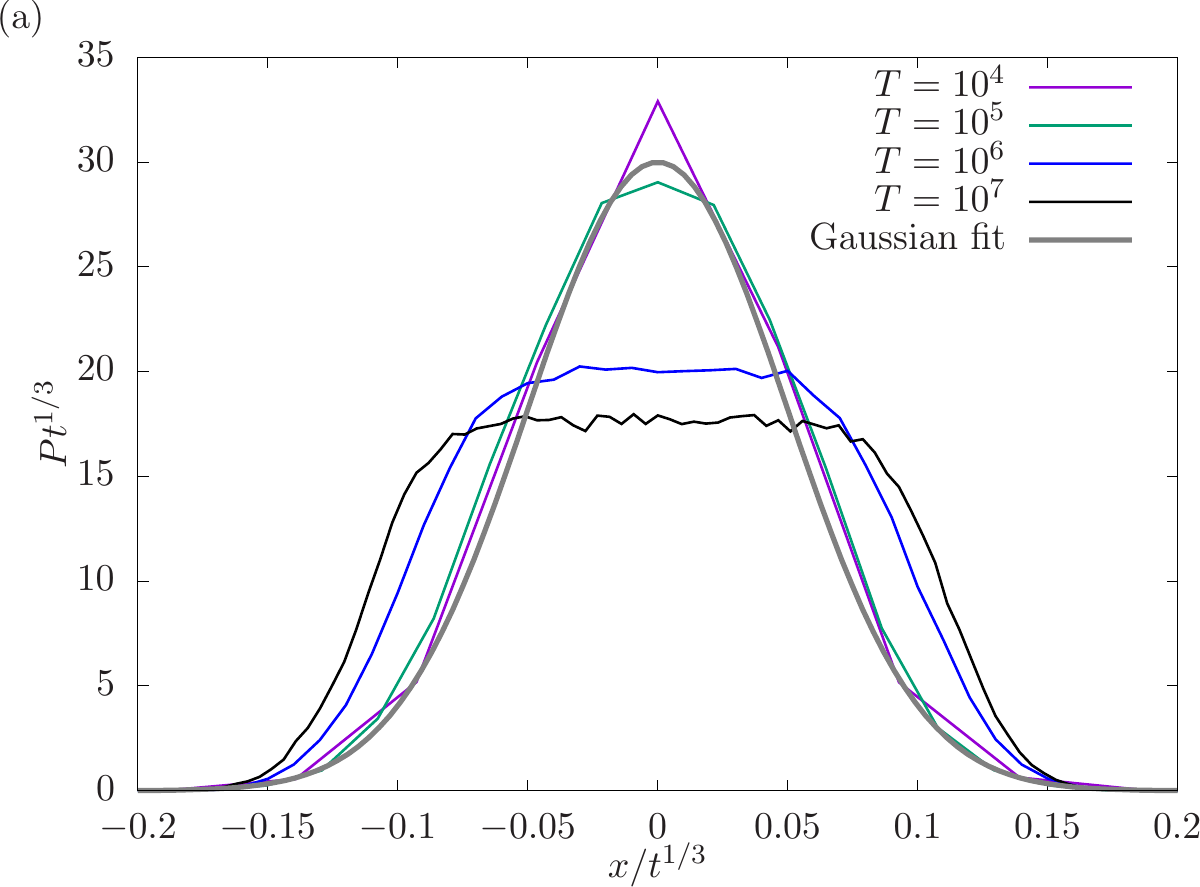}
  \includegraphics[width=8cm]{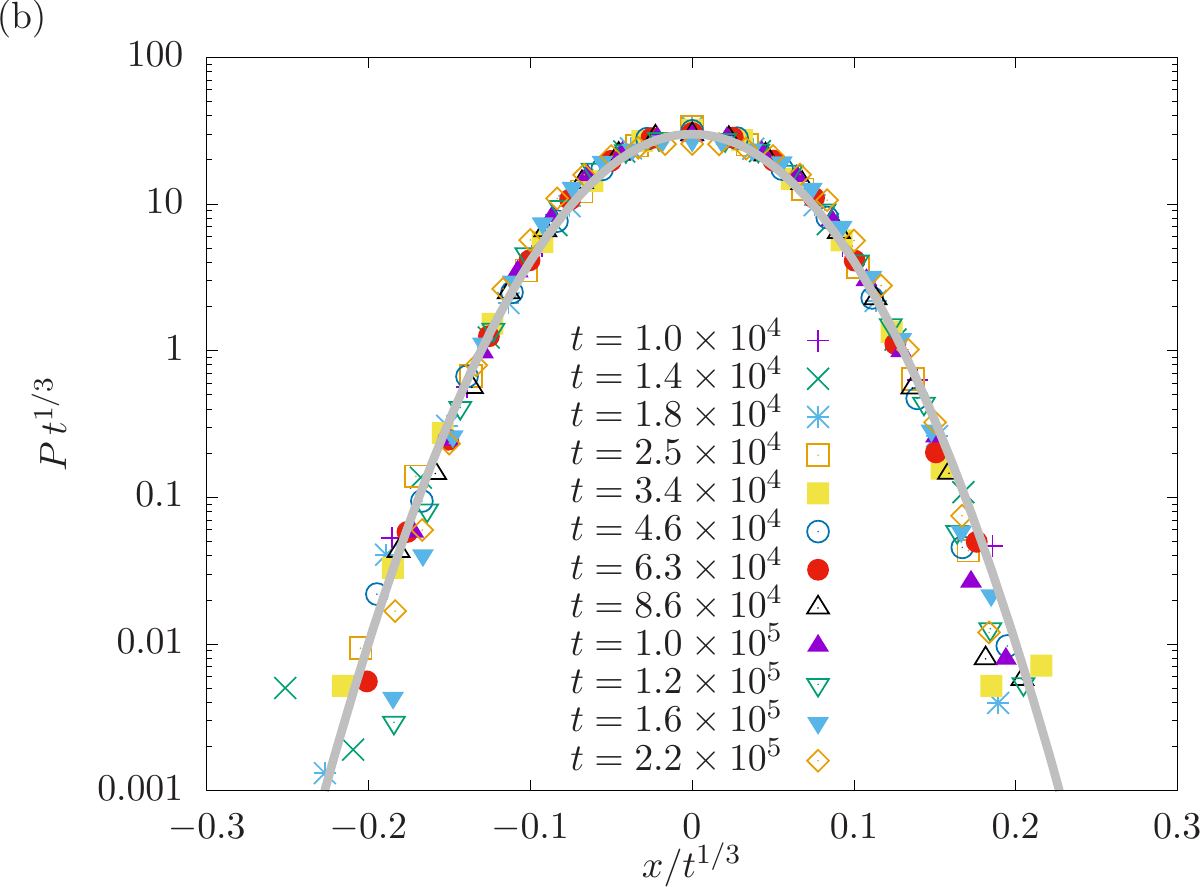}
  \includegraphics[width=8cm]{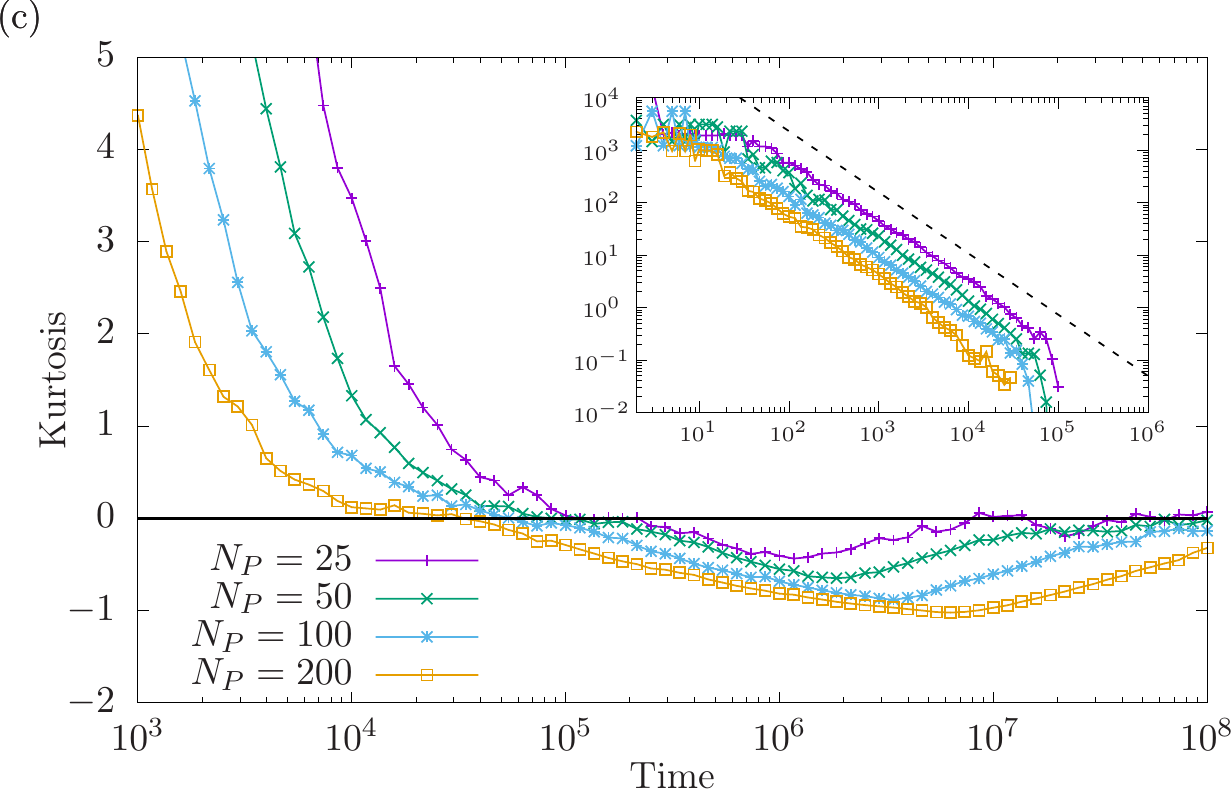}
  \caption{Rescaled particle histograms for different times using
    $N_P=200$ and $\J_0=10^{-6}$. (a) Histograms for the particle
    positions obtained for $N_S=2000$ samples, scaled using
    Eq. \eqref{eq:collapse}, as in Fig. \ref{fig:pme}. We can observe
    that short-term histograms collapse up to $t\sim 10^5$, when they
    start flattening. For even longer times, $t\sim 10^8$, the scaled
    histogram becomes closer to the initial collapse. (b) Scaled
    histogram data for 12 different times between $t=10^4$ and
    $t=2.2\cdot 10^5$, in logarithmic scale, showing that the collapse
    extends also to low probability values. The continuous line
    corresponds to a Gaussian distribution. (c) Time evolution of the
    (excess) kurtosis of the particle histograms for different values
    of $N_P$. Inset: log-log plot of the same data, showing a scaling
    law, Kurt $\sim t^{-7/6}$.}
  \label{fig:histog}
\end{figure}

The full histogram for the particle positions, considering all
$N_S=2000$ samples simultaneously, is provided in
Fig. \ref{fig:histog}, for different times using $N_P=200$. The
histograms are rescaled following Eq. \eqref{eq:collapse}, as in
Fig. \ref{fig:pme}. For short times, the histograms collapse,
increasing their width substantially between $t\sim 10^5$ and
$10^7$. The accuracy of the collapse can be appreciated in
Fig. \ref{fig:histog} (b), where we can see the rescaled histograms
for 20 times chosen between $t=10^4$ and $t=2\cdot 10^5$ in
logarithmic scale. Remarkably, also the low probabilities seem to
collapse around a Gaussian shape, as seen in the continuous
line. Interestingly, the scaling function predicted by the PME is not
a Gaussian but an inverted parabola (see Appendix \ref{sec:appendix}).

Further information about the histograms can be obtained by measuring
the (excess) kurtosis, defined as

\begin{equation}
  \text{Kurt}[x_{p,s}]={E_{SP}\left[ \( x_{p,s}-E_{SP}[x_{p,s}]\)^4
      \right] \over \(\sigma_{SP}[x_{p,s}]\)^4}  -3,
  \label{eq:defkurt}
\end{equation}
which is shown in Fig. \ref{fig:histog} (c). We can observe that, for
short times, the distributions are extremely {\em leptokurtic}
(peaked), with the kurtosis decreasing approximately as $\kappa \sim
t^{-7/6}$ for short times. Afterwards, it becomes negative
(platykurtic, square-like) and evolves slowly towards zero,
i.e. Gaussian-like.

\subsection{Active hoppings}

\begin{figure}
  \includegraphics[width=8cm]{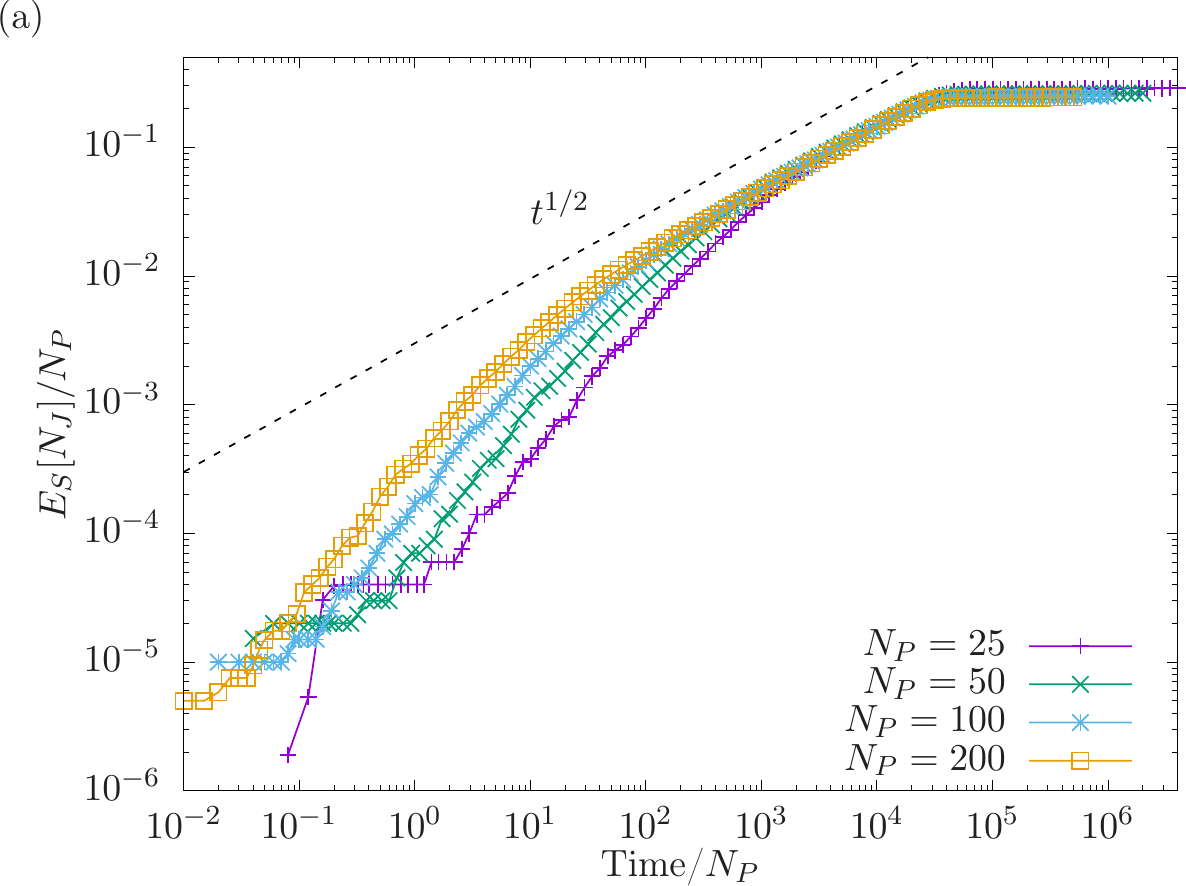}
  \includegraphics[width=8cm]{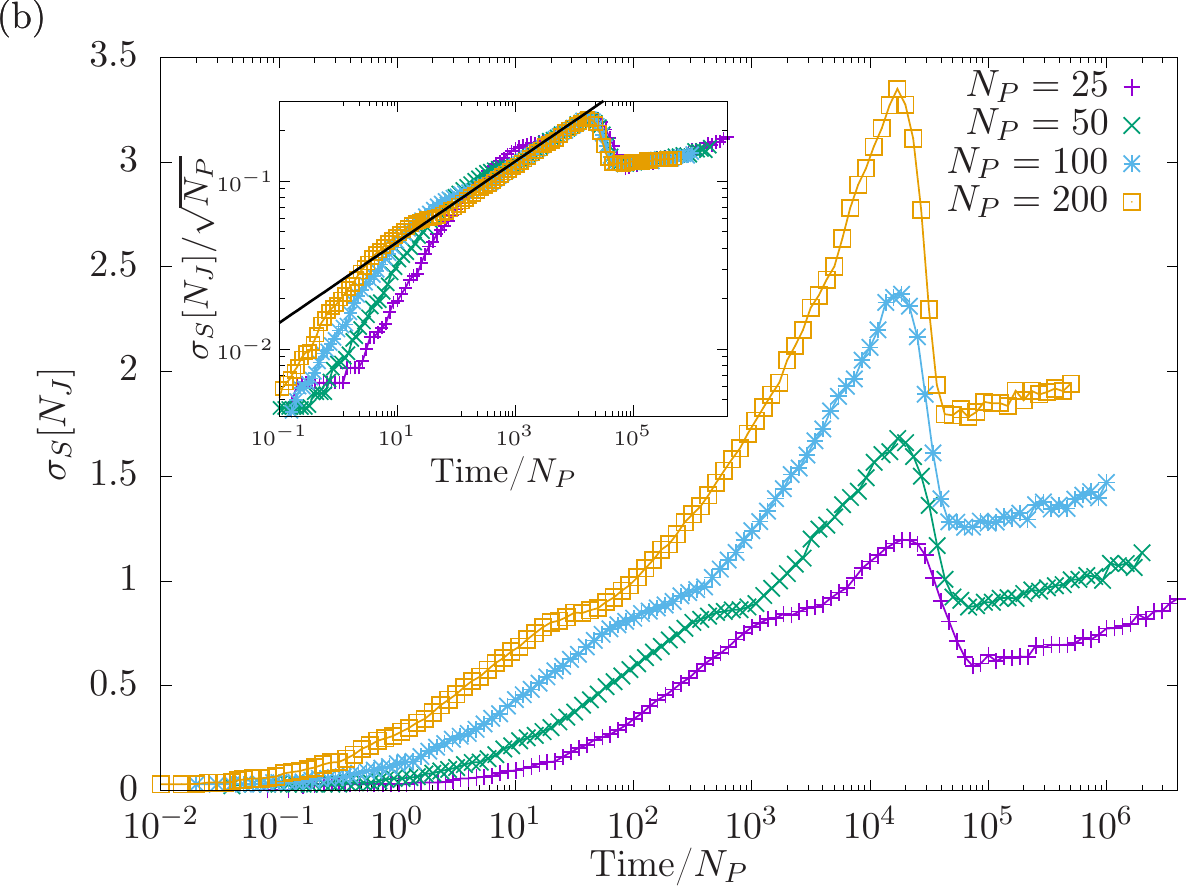}
  \caption{(a) Average number of active hopping probabilities as a
    function of time using $\J_0=10^{-6}$, both scaled with the number
    of particles $N_P$, compare to Fig. \ref{fig:width} (c). The
    continuous line corresponds to a $t^{1/2}$ scaling. (b)
    Sample-to-sample deviation of that magnitude. Inset: scaling with
    the square root of the number of particles makes the curves
    collapse around the peak. The continuous line represents a scaling
    of $t^{1/4}$.}
  \label{fig:jav}
\end{figure}

The size of the cloud can be accessed also through an analysis of the
set of hopping probabilities, $\{J_i\}_{i=1}^L$. Let us consider the
observable

\begin{equation}
  N_J={1\over \J_1-\J_0}\sum_{i=1}^L (J_i-\J_0).
  \label{eq:JT}
\end{equation}
If all $J_i$ are either $\J_0$ or $\J_1$, $N_J$ measures the number of
$\J_1$ values. Fig. \ref{fig:jav} (a) represents the expected value,
$E_S[N_J]$, scaled with the number of particles $N_P$, as a function
of time, also divided by $N_P$. We observe a similar behavior to
$W_\av(t)$, with a time range presenting $t^{1/2}$ scaling followed by
saturation. We may also ask about the fluctuations in the size of the
particle cloud, by measuring $\sigma_S[N_J]$, which appears in
Fig. \ref{fig:jav} (b). In this case, the curves for different $N_P$
do not collapse unless we scale differently the time axis (dividing by
$N_P$) and $\sigma_S[N_J]$, which should be divided by $N_P^{1/2}$, as
we show in the inset, in logarithmic scale. We observe a steady
increase of $\sigma_S[N_J]$ with time, presenting two different
scaling regimes: an initial $t^{1/2}$ followed by a $t^{1/4}$ regime,
before reaching a quick decay at the saturation time towards a
constant value. This behavior points at a possible {\em dynamical
  phase transition} at the saturation point. The physical reason for
this behavior will be clarified in the next section, after several
other observables have been considered.

\begin{figure*}
  \includegraphics[width=8cm]{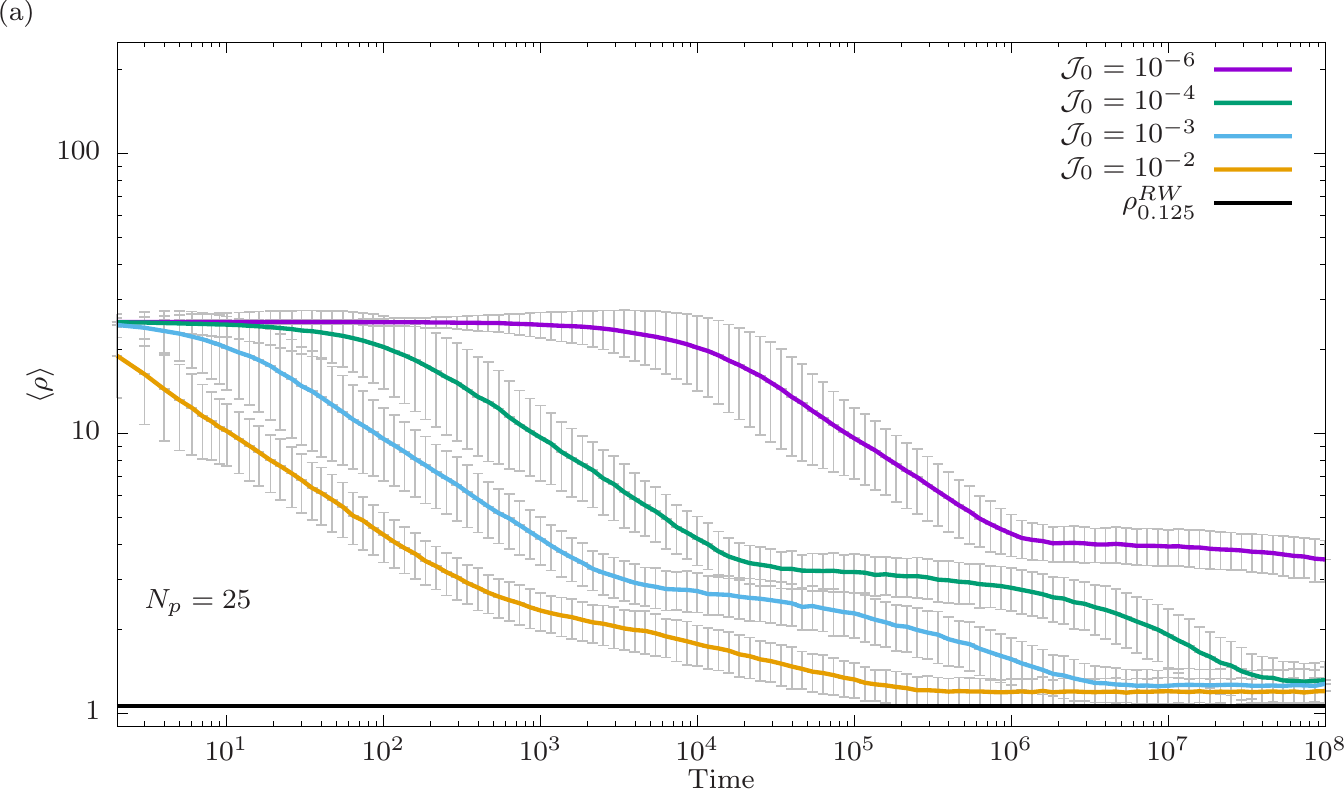}
  \includegraphics[width=8cm]{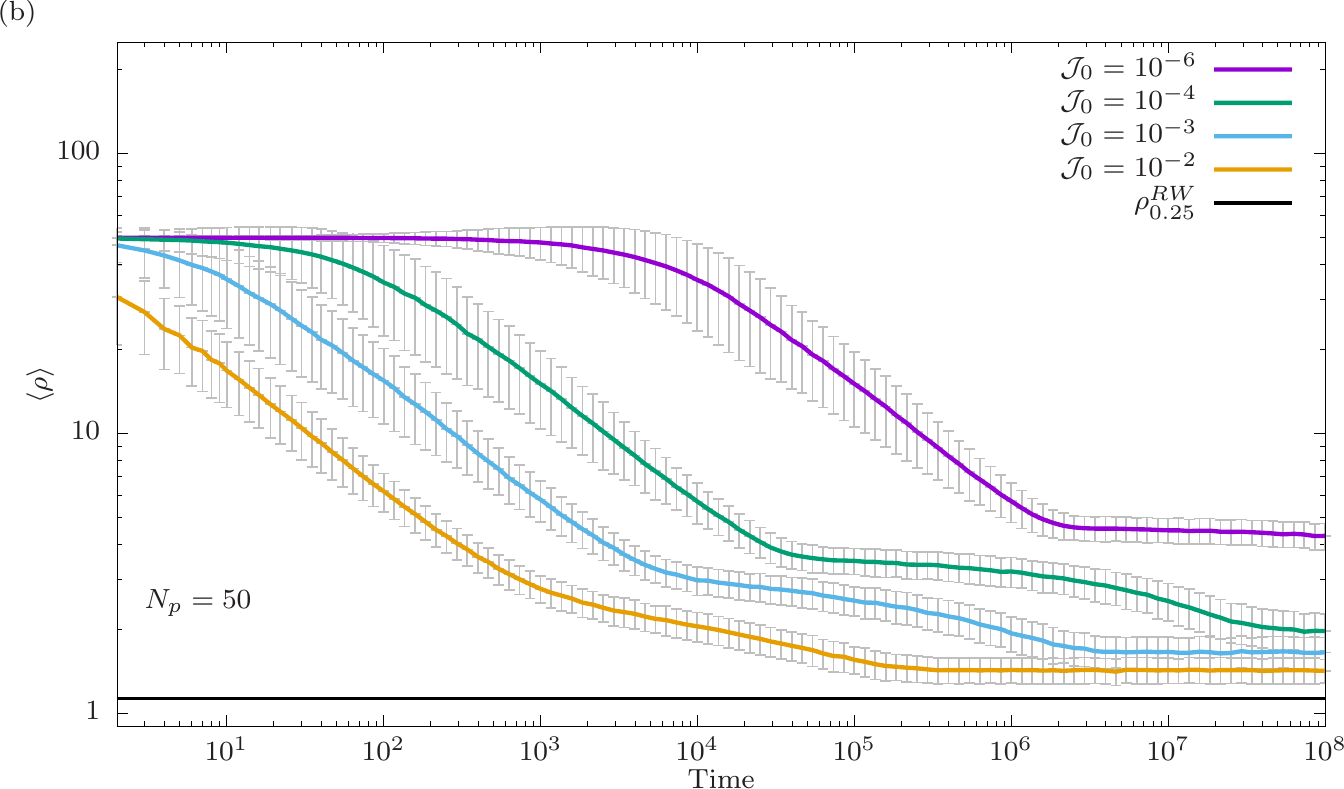}
  \includegraphics[width=8cm]{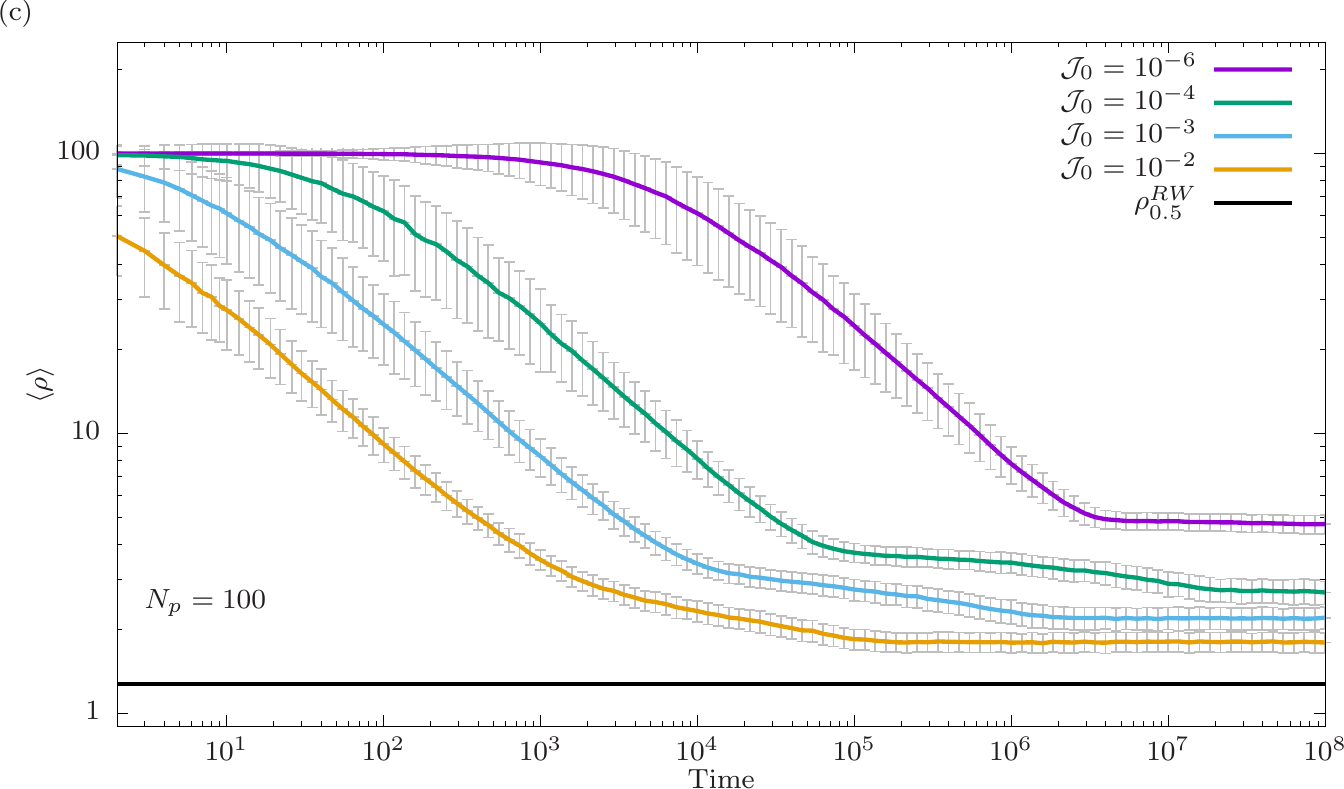}
  \includegraphics[width=8cm]{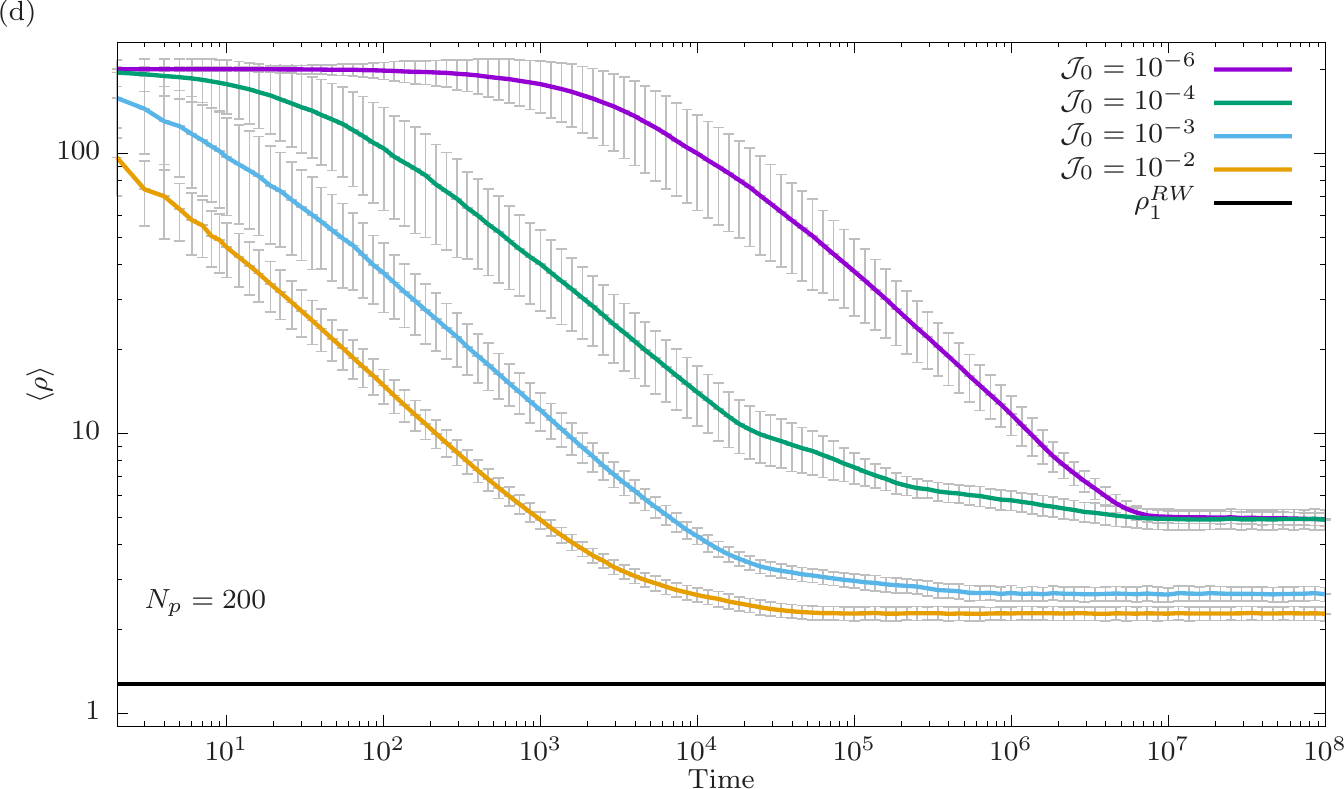}
  \includegraphics[width=8cm]{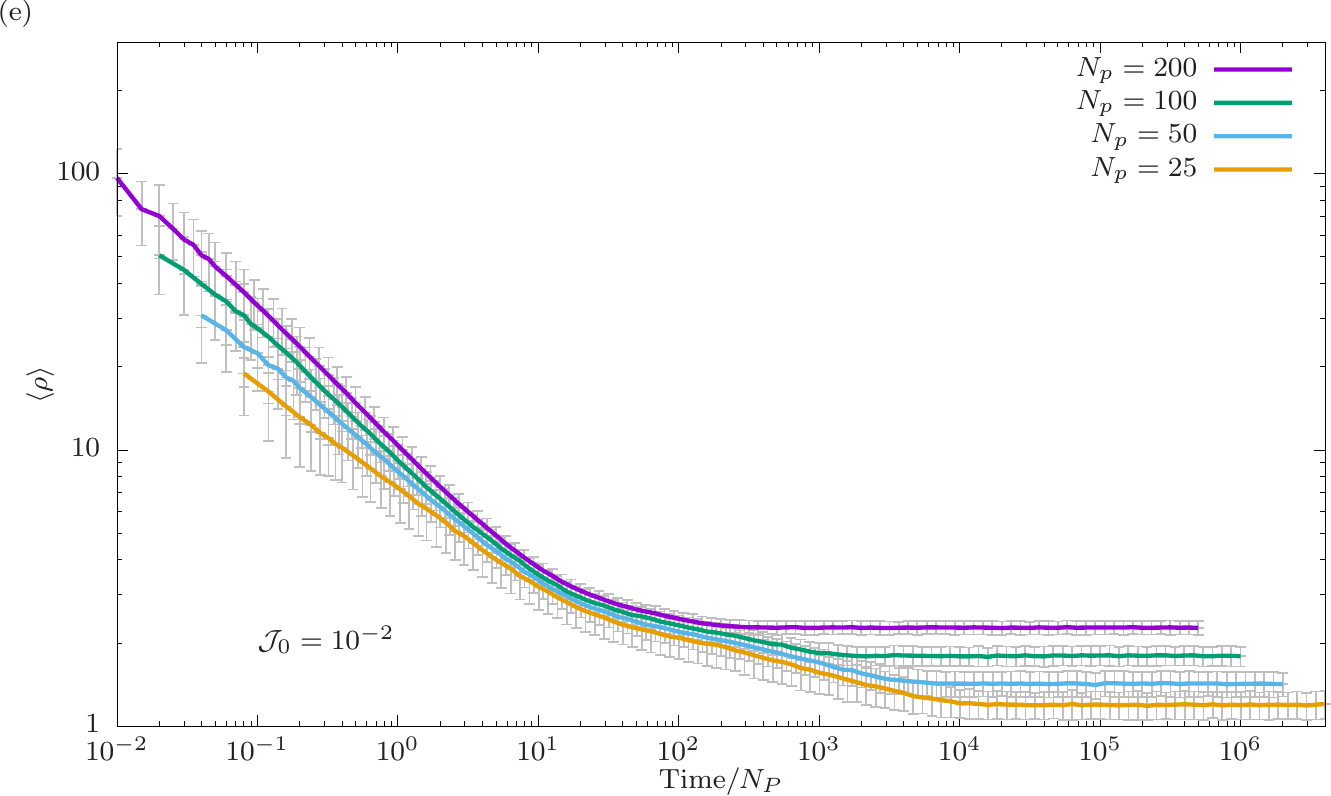}
  \includegraphics[width=8cm]{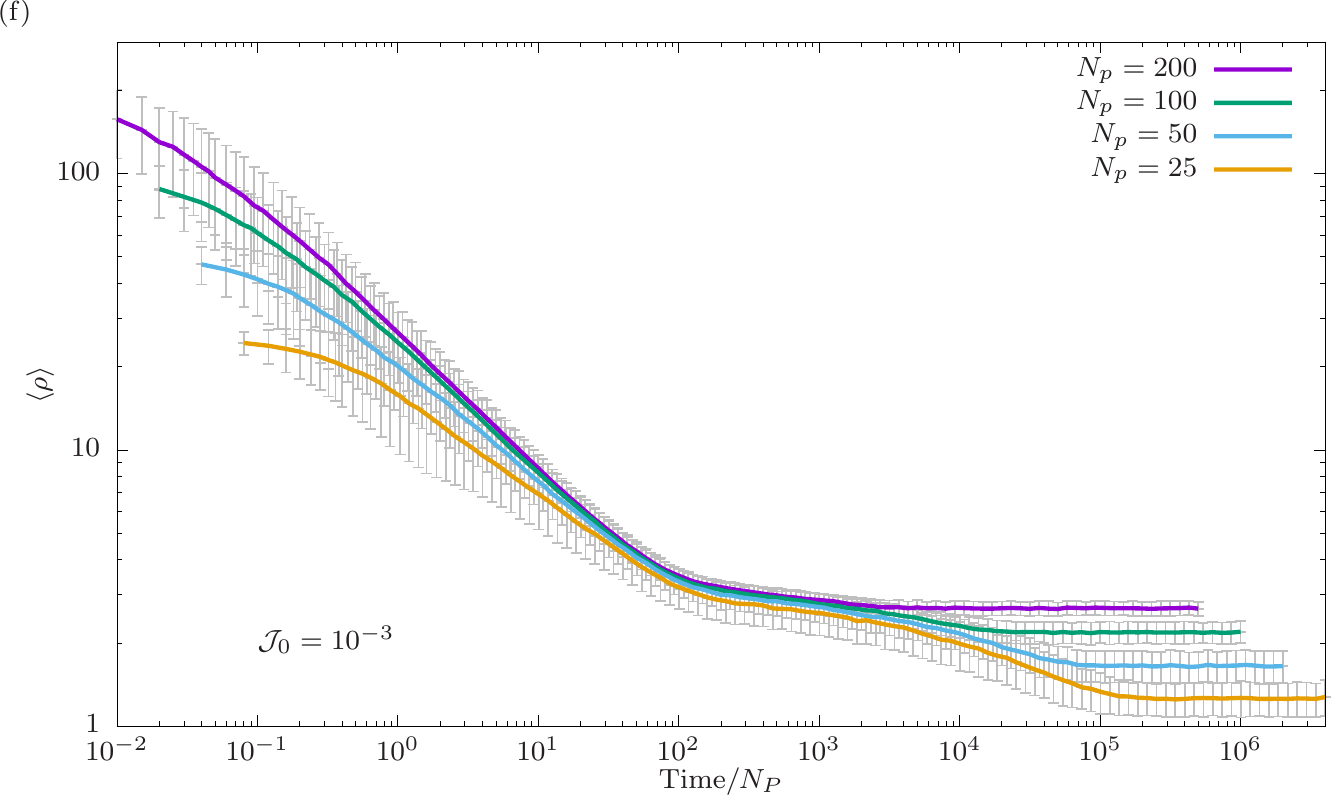}
  \includegraphics[width=8cm]{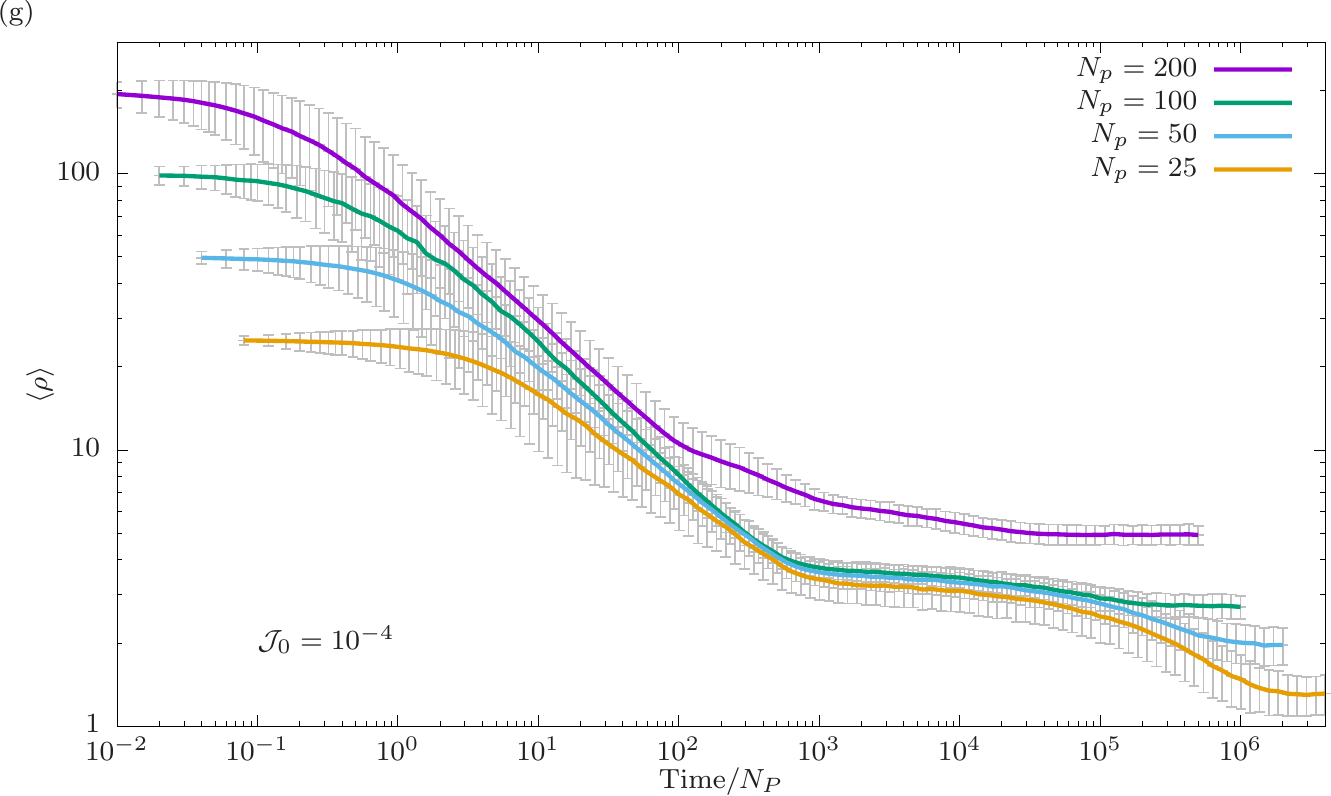}
  \includegraphics[width=8cm]{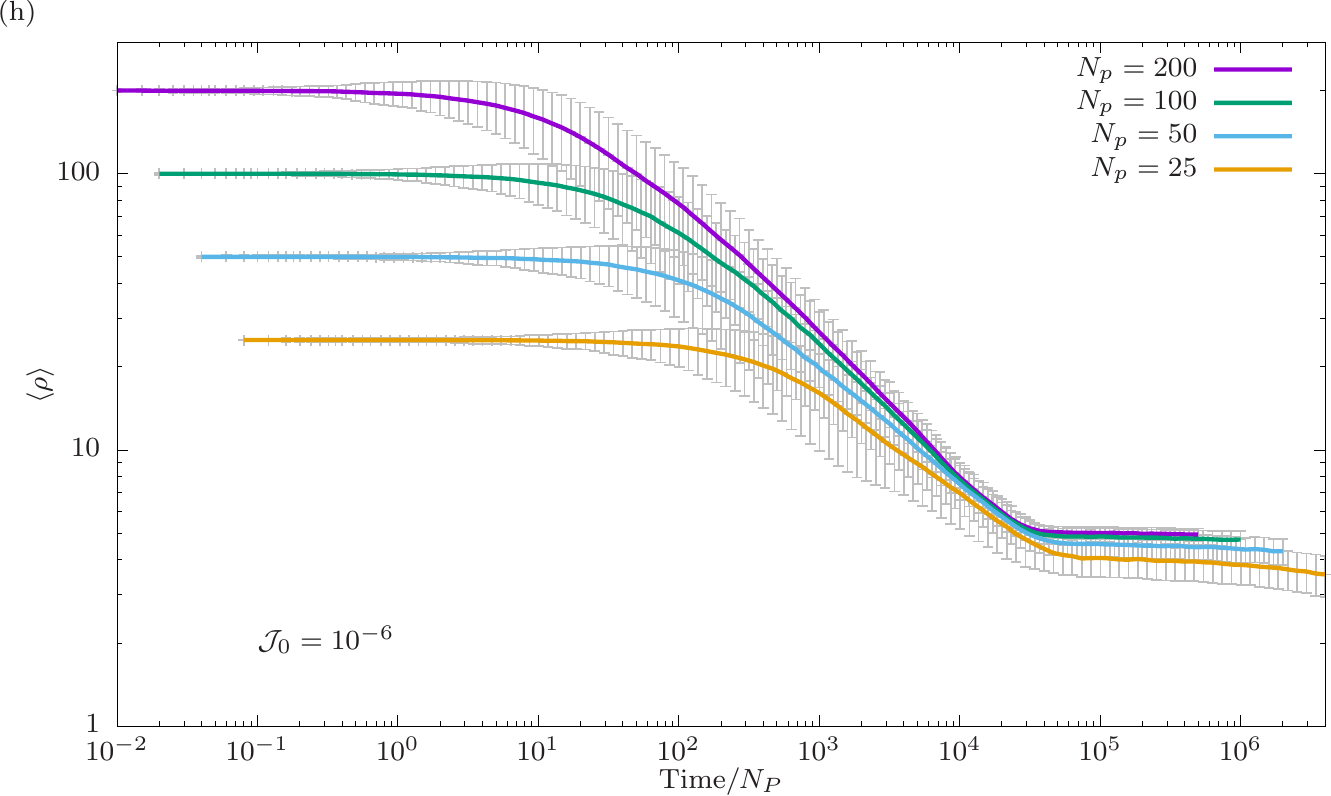}
  \caption{Average population-weighted density, $\<\rho\>$, as defined
    by Eq. \eqref{eq:renyi}, for (a) $N_P=25$, (b) $N_P=50$, (c)
    $N_P=100$ and (d) $N_P=200$ particles with different values of
    $\J_0$, and for (e) $\J_0=10^{-2}$, (f) $\J_0=10^{-3}$, (g)
    $\J_0=10^{-4}$ and (h) $\J_0=10^{-6}$ with different values of
    $N_P$. In the last four panels, time is divided by $N_P$. The
    minimum number of samples is $N_S=500$. Notice that the initial
    value is always $\<\rho(0)\>=N_P$. The horizontal bars in panels
    (a-d) correspond to the theoretical prediction for independent
    random walkers, Eq. \eqref{eq:rhoinfty}.}
  \label{fig:renyi}
\end{figure*}

\subsection{Population-averaged density}

A relevant measure of particle correlations can be borrowed from {\em
  demographics}, where an interesting distinction is made between the
average population density and the {\em experienced density}, also
called {\em population-weighted density}, defined as the average local
density experienced by a randomly selected individual
\cite{Ottensmann.18}. For a lattice system, it can be defined from the
local occupations $n_i$, i.e. the number of particles at site $i$,

\begin{equation}
\<\rho\>={1\over N_P} \sum_{i=1}^L n_i^2,
\label{eq:renyi}
\end{equation}
in other terms, it is related to the R\'enyi entropy of order
2. Notice that if the occupation is homogeneous, $n_i=N_P/L$ and
$\<\rho\>=N_P/L$. Fig. \ref{fig:renyi} shows the expected value of the
population-weighted density of this system for different numbers of
particles: (a) $N_P=25$, (b) $N_P=50$, (c) $N_P=100$ and (d)
$N_P=200$, using four values of $\J_0=10^{-6}$, $10^{-4}$, $10^{-3}$
and $10^{-2}$. The horizontal bar in all cases corresponds to the
theoretical prediction for independent random walkers for a density
$\lambda\equiv N_P/L$,

\begin{equation}
\rho^{RW}_\lambda={\lambda\over 1-\exp(-\lambda)},
\label{eq:rhoinfty}
\end{equation}
as it is proved in Appendix \ref{appendix:B}. For very short times,
$\<\rho\>=N_P$, because all particles are together at the origin. The
values of $\<\rho(t)\>$ decay and seem to stabilize at a finite value
$\rho_\infty$ in the $t\to\infty$ limit, which depends both on $\J_0$
and $N_P$. In all cases $\rho_\infty > \rho^{RW}_\lambda$, showing
that the particles present strong interactions.

Panels (e-h) of Fig. \ref{fig:renyi} show the population-weighted
density for (e) $\J_0=10^{-2}$, (f) $\J_0=10^{-3}$, (g) $\J_0=10^{-4}$
and (h) $\J_0=10^{-6}$ with the values of $N_P$ considered above,
i.e. they show the same data as panels (a-d) organized
differently. Moreover, time is divided by $N_P$ in all of them. We
observe that $\rho_\infty$ grows steadily with $N_P$ for
$\J_0=10^{-2}$, presents jumps with $N_P$ for $\J_0=10^{-3}$ and
$\J_0=10^{-4}$, and sticks to a value $\rho_\infty \approx 5$ for all
particle numbers when $\J_0=10^{-6}$. This value can be understood
through inspection of Fig. \ref{fig:bruja} (a), where we see that the
cloud occupies a region of size $L_0\approx 40$ sites. If the density
is homogeneous in this region, the experienced density will be of
order $N_P/L_0\approx 5$. This result is also compatible with the
observed cloud size estimated through $W_\av$ and $N_J$ which can be
read from Fig. \ref{fig:width} and \ref{fig:jav}. Interestingly, for
$\J_0=10^{-4}$ the system reaches the same final density, even though
the width is much larger. The reason is that {\em the cloud breaks
  down into clusters} with the same final density, $\approx 5$. For
$\J_0=10^{-2}$, on the other hand, the final density
$\rho_\infty\approx 2.3$, still substantially above the value for
independent random walkers, which is $\rho_1^{RW}\approx 1.582$ for
this density $\lambda=N_P/L=1$ according to Eq. \eqref{eq:rhoinfty}.

It is also interesting to observe the different stages shown in panels
(e-h) of Fig. \ref{fig:renyi}. We can distinguish a first initial
stage, for which the cloud remains concentrated near a single
point. This is followed by a decay stage, which corresponds to the
expanding cloud and, for low $\J_0$, the PME phase. This stage ends
in a plateau when the cloud is able to break down, at a time
corresponding to the saturation of $W_\av$ in Fig. \ref{fig:width} (c)
and the peak of $\sigma_S[N_J]$ in Fig. \ref{fig:jav} (b). The reader
is also referred to Fig. \ref{fig:bruja}, to check the typical
configurations in all the cases. The plateau may still give rise to
further discrete decays for intermediate values of $\J_0$, presumably
when the island structure changes.

\subsection{Center of mass and correlations}

The sub-diffusive expansion of the cloud signals an effective
attraction between the particles, which is associated to strong
correlations between their positions. We will characterize these
correlations using the fluctuations of the center of mass (CM) as a
proxy. Indeed, for every time and sample we can define

\begin{equation}
  X_{CM,s}(t)\equiv E_P[x_{p,s}(t)],
\end{equation}
and, along with it, the sample-to-sample fluctuations of this magnitude,

\begin{equation}
  \sigma_{CM}(t)\equiv \sigma_S[E_P[x_{p,s}(t)]],
  \label{eq:CM}
\end{equation}
which are strongly tied to the difference between $W_{\text{all}}$ and
$W_{\text{av}}$ expressed in inequality \eqref{eq:ineq}. Indeed,

\begin{equation}
  W^2_{\text{all}}-W^2_{\text{av}}=E_S[E_P[x_{p,s}]^2]-E_S[E_P[x_{p,s}]]^2
= \sigma^2_{CM},
    \label{eq:wcm}
\end{equation}
i.e. they correspond to the distance between both curves in
Fig. \ref{fig:width} (a), see inequality \eqref{eq:ineq}. Moreover,
it can also be proved that

\begin{align}
  \sigma^2_{CM} & ={1\over N_P^2}\sum_{p,q=1}^{N_P}
  \(E_S[x_{p,s}x_{q,s}]-E_S[x_{p,s}]E_S[x_{q,s}]\) \nonumber \\
  & \equiv {1\over N_P^2} \sum_{p,q} C_{p,q},
  \label{eq:cmcorr}
\end{align}
where we have defined $C_{p,q}$ as the average over samples of the
correlation between particles $p$ and $q$, and we have dropped the
dependence on $t$ for convenience. Given the symmetry between
particles, we must have
$C_{p,q}=C_0\delta_{p,q}+C_1(1-\delta_{p,q})$. For uncorrelated random
walkers $C_1=0$ and $C_0=Dt$, so we have

\begin{equation}
  \sigma^2_{CM}={Dt \over N_P},
  \label{eq:cm_rw}
\end{equation}
as expected. Yet, if each particle moves diffusively but correlations
are as strong as possible, $C_1=C_0=Dt$, we have

\begin{equation}
  \sigma^2_{CM}=Dt.
  \label{eq:cm_corr}
\end{equation}

\begin{figure}
  \includegraphics[width=8cm]{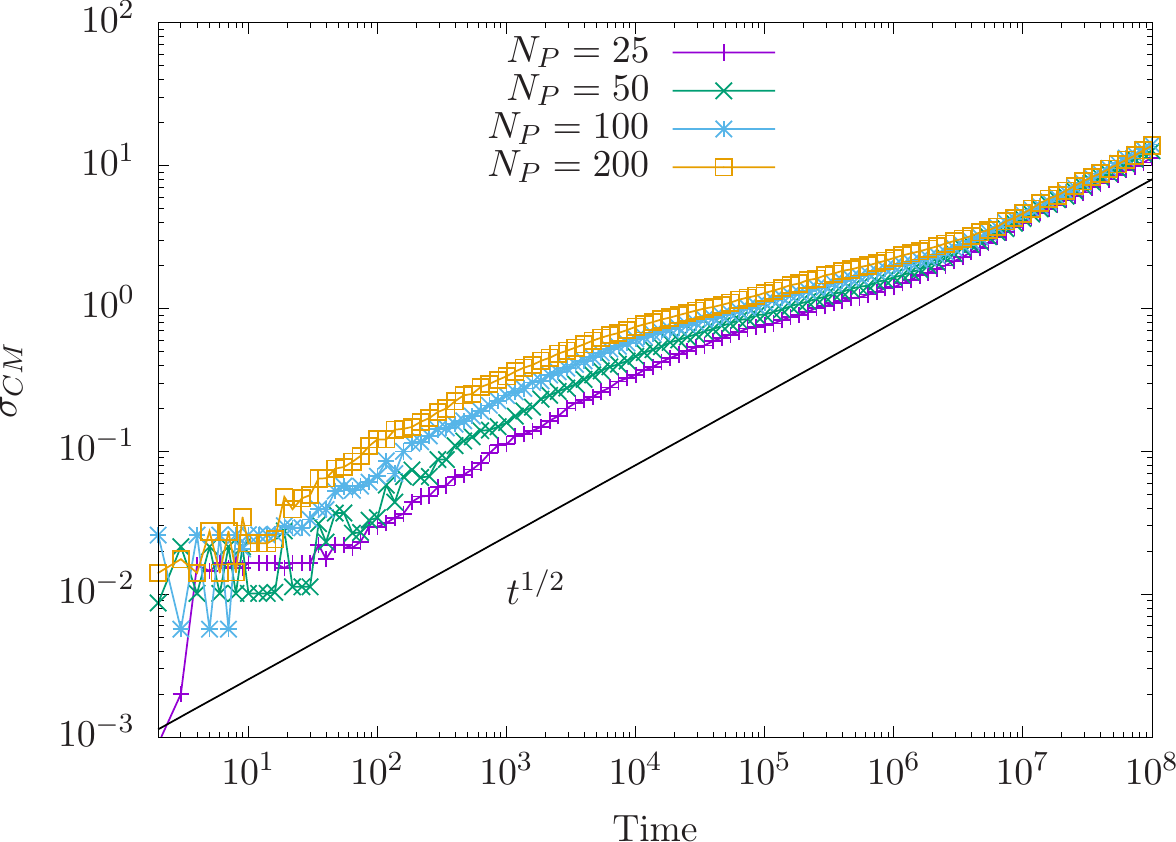}
  \includegraphics[width=8cm]{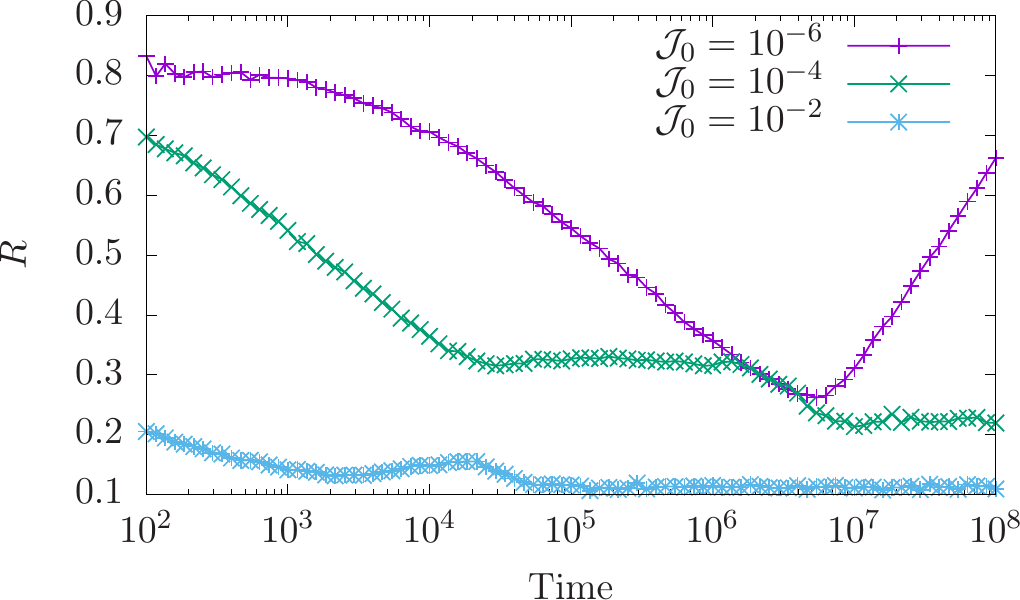}
  \caption{(a) Fluctuations of the center of mass of the particle
    cloud, for different numbers of particles using
    $\J_0=10^{-6}$. The straight line represent a $t^{1/2}$
    growth. (b) Correlation ratio, $R$, defined in
    Eq. \eqref{eq:ratio}, as a function of time (in log scale), using
    $N_P=200$ and different values of $\J_0=10^{-6}$, $10^{-4}$ and
    $10^{-2}$.}
  \label{fig:cm}
\end{figure}

The actual behavior of $\sigma_{CM}(t)$ is shown in Fig. \ref{fig:cm}
(a) for different numbers of particles as a function of time. We
observe an intermediate regime where the fluctuations of the center of
mass grow slowly, followed by a long-term diffusive scaling with
$t^{1/2}$. But the most relevant observation is that, in this 
regime, the fluctuations of the center of mass do not depend on the
number of particles, thus showing a behavior similar to
Eq. \eqref{eq:cm_corr}. Each particle follows a random walk, but all
particles are strongly correlated and the center of mass provides
evidence for this.

Indeed, we can define a correlation ratio,

\begin{equation}
  R\equiv {\sigma_{CM}\over W_{\text{all}}} \in [0,1],
  \label{eq:ratio}
\end{equation}
which is zero for a large number of uncorrelated particles, and tends
to one for strongly correlated walkers. Fig. \ref{fig:cm} (b) shows
the value of this ratio as a function of time for $N_P=200$ and
different values of $\J_0=10^{-6}$, $10^{-4}$ and $10^{-2}$. Indeed,
we observe that the correlation ratio is very low for large values of
$\J_0$, which corresponds to independent random walkers. Yet, for very
low $\J_0\approx 10^{-6}$ its behavior is more intriguing. It starts
at a high value, decreasing logarithmically as the particle cloud
expands until it reaches a minimum value, and starts a new logarithmic
growth stage when the cloud has reached maximal expansion.


\section{Physical picture}
\label{sec:picture}

The numerical data described in the previous section allow us to
provide a physical picture for the statistical properties of the cloud
of random walkers and the metric field in the RWDM model, based on the
strong correlations between the positions of the particles, signalled
by several observables, such as the difference between $W_\all$ and
$W_\av$, the long term behavior of the population-averaged density
$\<\rho\>$ or the correlation ratio $R$.

Evolution starts with all the particles concentrated at the center
waiting to hop on a $\J_0$ link, which will take a time $\sim
(N_P\J_0)^{-1}$. The first particle taking such a hop will make other
particles follow, and the particle cloud as a whole will start hopping
between two neighboring sites for a while, until a second particle
hops on a new $\J_0$ link again, thus spreading the cloud. Notice that
different samples will make different choices during these initial
steps. Thus, the {\em average width} of each cloud becomes much
smaller than the {\em global width} of the ensemble of all
clouds. Yet, both magnitudes grow at different rates: each cloud grows
diffusively, with $t^{1/2}$, see Fig. \ref{fig:width}, while the
global cloud grows subdiffusively, approximately with $t^{1/3}$, which
corresponds to the global behavior of the solution of the PME, see
Eq. \eqref{eq:pme} and \eqref{eq:collapse}. Thus, we find a reasonable
fit between our theoretical and numerical models for a certain time
range. Yet, we ought not to forget that our model is described by a
highly non-linear partial differential equation, and therefore its
validity is arguably limited to certain time regions.

As the cloud expands, internal links are close to their maximum
possible value, $\J_1$. Thus, our estimate of the number of active
links, $N_J$, defined as in \eqref{eq:JT} and shown in
Fig. \ref{fig:jav}, presents the same power-law scaling as the average
width of the clouds, $W_\av$, as shown in the central panel of
Fig. \ref{fig:width}. Yet, the clouds can not expand without limit,
because the average width grows faster than the global
width. Moreover, the local density will decay, as shown in
Fig. \ref{fig:renyi}, and when it reaches a minimal value
$\rho_\infty$, the cloud will either stop its expansion or break up
into islands. The maximum size of the clouds is, therefore, an effect
related to the finite number of particles, and scales linearly with
$N_P$, as we can see in the saturation value of both $W_\av$ and
$N_J$. Notice that this saturation is not related to the finite size
of the system. Furthermore, beyond this saturation there is a finite
probability that some links will not be occupied for a time $>t_0$,
thus allowing them to return to their low value, $\J_0$, as we can
check in Fig. \ref{fig:bruja} (a-b), where we can observe how it
separates into two or more islands. Of course, the continuum
approximation provided by the PME is not valid beyond the saturation
point.

At this point we observe a signature of a dynamical phase transition
in the sudden fall of the deviation of the number of active hoppings,
$\sigma_S[N_J]$, shown in Fig. \ref{fig:jav} (b). We conjecture that
this fall can be explained as follows. The clouds grow unbroken before
saturation, with an increasing sample-to-sample variance. As they
reach their asymptotical density they stop growing and they may break
up into smaller fragments which do not present a tendency to
expand. Saturated clouds, whether broken or unbroken, present a nearly
fixed size, thus explaining the drop in the variance of $N_J$. Yet,
the fragments can still wander, and $W_\av$ can still grow. Yet, when
the cloud remains unbroken $W_\av$ saturates too.

Moreover, the ensemble formed by all the clouds will continue its
expansion. Indeed, the sample-to-sample fluctuations of the CM of the
clouds grow diffusively, providing evidence for our claim. Yet, these
fluctuations of the CM provide a proxy measurement for the particle
correlations, through Eq. \eqref{eq:cmcorr}. For large times and low
$\J_0$, the fluctuations of the CM coincide for different numbers of
particles, see Fig. \ref{fig:cm} (a), providing evidence for strong
internal correlations. Indeed, the correlation ratio $R$, defined in
Eq. \eqref{eq:ratio}, takes much larger values for low $\J_0$, showing
that a large fraction of the fluctuations of the cloud can be
accounted for by the fluctuations of the center of mass, while the
cloud behaves internally as a rigid cluster. 

In the long term we conjecture that the system will consist of one or
several clusters which will wander through the system. If the system
size is infinite, they will eventually drift away from each
other. Otherwise, they will collide into each other with a certain
frequency, giving rise to a steady state in which clusters split up
and merge together. These final stages present some similarities with
the long-term behavior of galaxies in an expanding or steady-state
universe.


\section{Conclusions and further work}
\label{sec:conclusions}

We have considered a model of {\em random walkers on a deformable
  medium} (RWDM), in which random walkers interact with the medium on
which they move. Indeed, the hopping probability associated to each
link becomes a dynamical variable, which gets higher when it is
actually used, or relaxes towards a lower equilibrium value
otherwise. The reinforcement of actually used paths presents
similarities with neural plasticity: {\em neurons that fire together
  wire together}, also known as Hebb's rule \cite{hertz.90}. Adaptable
networks appear in a variety of settings, where usage of a link tends
to reinforce its strength and, thus, the probability of further usage
\cite{Gross,Sayama}. Moreover, the RWDM can also serve as a toy model
of stochastic gravity since the medium can be considered as a (1+1)D
metric, following the principle that matter tells geometry how to
curve and geometry tells matter how to move.

The RWDM can be considered as a part of the family of {\em reinforced
  random walks}, with several relevant differences. Indeed, reinforced
random walks enhance the probability of re-visiting nodes or edges,
either permanently or temporarily. The RWDM extends these ideas to
allow the jumping probabilities to become a full-fledged dynamical
object.

For the sake of concreteness, we have focused on the 1D case, starting
with an initial condition in which all particles are concentrated at
the center, and the medium starts relaxed, in the limit in which the
excited hopping is much larger than the equilibrium one, and for fast
relaxation. In that limit, a continuum description should be provided
by the {\em porous medium equation} (PME), which presents scaling
solutions in this regime: our cloud should expand subdiffusively with
time, due to self-localization effects, and grows like $\sim t^{1/3}$.

Indeed, this subdiffusive growth is observed in numerical experiments,
but only when we consider the global cloud consisting of all samples,
i.e. an average over the whole configuration space. On the other hand,
when we consider the average over samples of the width of {\em each}
cloud, we typically observe a diffusive behavior, $t^{1/2}$. Yet, the
global width must necessarily exceed the average width, and their
difference is related to the fluctuations of the center of mass of
each cloud, which can also be linked to the internal correlations
within the cloud. Thus, we were able to identify a growth regime and a
saturated regime (not related to the system size), where the number of
active links reaches a final asymptotic value, even though the global
width can still grow because the center of mass of the cloud can still
wander.

Statistical mechanics of the RWDM seems to provide a wealth of
behaviors which are worth exploring, such as the extension to several
dimensions or the spread of different initial configurations, which
can give rise to structure formation. Also, the internal dynamics of
the metric field can be extended in several interesting ways, such as
introducing a {\em surface tension} which would allow for the
propagation of waves. It is interesting to remark that the porous
medium equation ruling the spread of the RWDM cloud is {\em causal},
while the heat equation is not. This fact might bear relevance for the
study of relativistic brownian motion, which is still an open problem
\cite{kostadt.13}. Moreover, critical behavior on a gravity field has
been extensively studied using Liouville theory, giving rise to the
well-known Knizhnik-Polyakov-Zamolodchikov (KPZ) equation which
evaluates how the critical exponents of a 2D conformal field theory
(CFT) deform in presence of gravity \cite{kpz.88,ambjorn.96}. We
intend to explore all these connections in the forthcoming future.


\begin{acknowledgments}
We would like to thank Rodolfo Cuerno for his help and valuable
insights. Also, we acknowledge the Spanish government for financial
support through grants PGC2018-094763-B-I00 and PID2019-105182GB-I00.
\end{acknowledgments}


\appendix

\section{Solution of the porous medium equation}
\label{sec:appendix}

A general solution for the porous medium equation \eqref{eq:pme} with
arbitrary exponent and a concentrated initial condition can be found
in \cite{Pattle.59,vazquez.91}. Here we will provide a simplified
version for the benefit of the reader, giving rise to the exact form
of $F(\eta)$, with $\eta\equiv xt^{-1/3}$, in Eq. \eqref{eq:collapse}.

Symmetry and scaling considerations lead to the following proposal,

\begin{equation}
  P(x,t) = \begin{cases} \frac{t^{-1/3}}{\alpha}
    \left[1-\(\frac{x}{\beta t^{1/3}}\)^2\right], & \text{if } |x|\leq
    \beta t^{1/3} \\ 0, & \text{otherwise}.
      \end{cases}
  \label{eq:proposedsolution}
\end{equation}
which corresponds to

\begin{equation}
  F(\eta)=\max(0,1-\eta^2).
  \label{eq:F}
\end{equation}

Normalization as a probability distribution leads to the condition

\begin{equation}
  {\alpha\over\beta}={4\over 3}.
\end{equation}

Plugging expression \eqref{eq:proposedsolution} into
Eq. \eqref{eq:pme} we obtain

\begin{equation}
  \alpha\beta^2=6\J_1.
\end{equation}

Thus, the values of $\alpha$ and $\beta$ are completely determined.

\section{Population-averaged density for independent random walkers}
\label{appendix:B}

In the case of random walkers, the expected value of the
population-averaged density can be obtained exactly through the
following argument. Let us consider $N_P$ particles on $L$ sites. The
probability that any of the sites will be empty is given by

\begin{equation}
p_E=\({L-1\over L}\)^{N_P},
\end{equation}
and thus the probability of a particle being occupied is just
$p_O=1-p_E$. The expected number of occupied sites is thus $Lp_O$, and
the population-averaged density corresponds to the expected density
within that set of occupied sites, $\rho_\infty=N_P/(Lp_O)$. When
$N_P/L=\lambda$ we have

\begin{equation}
\rho(L)_\infty^{RW} = {\lambda\over 1-\(1-{1\over L}\)^{N_p}},
\end{equation}
and in the thermodynamic limit we obtain

\begin{equation}
  \rho_\infty^{RW}=\lim_{L\to\infty} {\lambda\over 1-\(1-{1\over L}\)^{N_p}} =
      {\lambda\over 1-e^{-\lambda}}.
\end{equation}



\begin{thebibliography}{10}
\bibitem{note_erw} So-called because elephants are believed to have
  long term memory.
\end{thebibliography}
\end{document}